\documentclass[draftcls,onecolumn,12pt]{IEEEtran}  
\IEEEoverridecommandlockouts
\usepackage{amsmath,amssymb,amsthm,mathrsfs,amsfonts,dsfont,stmaryrd}
\usepackage{graphicx,color}  
\allowdisplaybreaks[4]
\newtheorem{theorem}{Theorem}
\newtheorem{definition}{Definition}
\newtheorem{lemma}{Lemma}  
\newtheorem{corollary}{Corollary} 
\newtheorem{assumption}{Assumption} 
\usepackage{enumerate}   
\newenvironment{remark}[1][Remarks]{\begin{trivlist}
\item[\hskip \labelsep {\bfseries #1}]}{\end{trivlist}}
\usepackage{enumerate}  
\IEEEoverridecommandlockouts 
\newcommand{\typ}[1]{T_\delta^n(#1)}   
\newcommand{\mkv}{-\!\!\!\!\minuso\!\!\!\!-}

\newcommand{\bN}{{\mathbb N}}
\newcommand{\cX}{{\mathcal X}}
\newcommand{\cY}{{\mathcal Y}}
 
\newcommand{\cM}{{\mathcal M}}   

\newcommand{\toas}[1]{\xrightarrow[#1]{}}

\graphicspath{{figs/}}
  \begin{document}
  \title{Capacity Results for the Multicast \\[-3mm] Cognitive Interference Channel}
\author{Meryem~Benammar,~\IEEEmembership{Member,~IEEE}, Pablo~Piantanida,\\~\IEEEmembership{Senior Member,~IEEE}, and~Shlomo~Shamai~(Shitz),~\IEEEmembership{Fellow Member,~IEEE}\\

\thanks{The material in this paper was partially published in the IEEE Information Theory Workshop, Jerusalem, April 26 to May 1, 2015 and and in the IEEE International Symposium on Information Theory, Hong Kong, June 14-19, 2015. This research was partially supported by the FP7 Network of Excellence in Wireless COMmunications NEWCOM\#.}
\thanks{Meryem~Benammar is with the Mathematical and Algorithmic Sciences Lab, France Research Center, Huawei Technologies Co., Ltd (e-mail: meryem.benammar@huawei.com). }
\thanks{Pablo~Piantanida is with the Laboratoire des Signaux et Syst\`emes (L2S UMR 8506) at CentraleSupelec-CNRS-Universit\'e Paris-Sud, France (e-mail:  pablo.piantanida@centralesupelec.fr).}  
\thanks{Shlomo Shamai (Shitz) is with the Department of Electrical Engineering, Technion, Haifa, Israel, (e-mail: sshlomo@ee.technion.ac.il).}
 }
\maketitle 
  
  \vspace{-4mm}
\begin{abstract}
 The capacity region of the \emph{Multicast Cognitive Interference Channel} (CIFC) is investigated. This channel consists of two independent  transmitters that wish to \emph{multicast} two different messages, each of them to a different set of users. In addition, one of the transmitters --commonly referred to as the cognitive transmitter-- has prior non-causal knowledge of both messages to be transmitted. This scenario combines difficulties and challenges arising in the \emph{Interference Channel}, the \emph{Broadcast Channel} and \emph{multicasting communications}. Our aim concerns the derivation of optimal interference mitigation techniques in such a challenging communication setup. We investigate to this end the \emph{multi-primary CIFC} and its dual \emph{multi-secondary CIFC} under various interference regimes as an attempt to build a thorough understanding for the more general setting. It is shown that, for some regimes, well-known coding techniques for the conventional CIFC remain still optimal in the presence of multicasting. While in other regimes, evolved encoding and/or decoding strategies are crucial. A careful use of these coding schemes and new outer bounding techniques allows us to characterize the capacity region for several classes of discrete memoryless and Gaussian channels in different interference regimes.   
 \end{abstract}
  
 %
 %


\newpage
 
\section{Introduction}
\begin{figure}[ht!]
 \centering
 \includegraphics[scale=0.5]{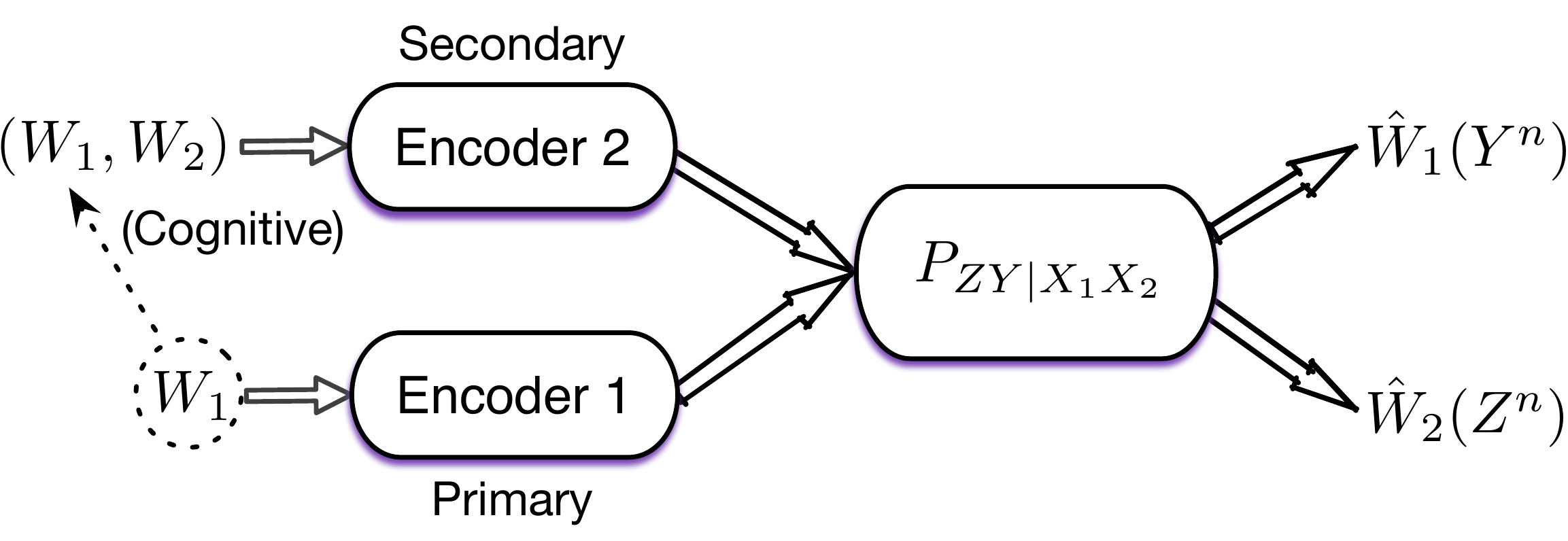} 
 \caption{The Cognitive Interference Channel.}
 \label{CIC}  
 \end{figure}

The Cognitive Interference Channel (CIFC) was first introduced by Devroye \textit{et al.}~\cite{Devroye} as an interference channel with two sources and two destinations in which one of the sources has full non causal knowledge of both messages to be transmitted, as depicted in Fig.~\ref{CIC}. The cognitive source models the secondary transmitter of a cognitive radio environment, which, upon sensing the primary transmitter's message, communicates the secondary message to the secondary user. As such, the secondary source should not create too much interference in the secondary transmission so as not to cause impediment to the primary communication, however, it can also cooperate with the primary source and thus enhance the performances of the primary communication. As it is defined, a CIFC can be regarded as a Broadcast Channel with a helper, i.e., the primary encoder. The helper enhances the transmission of the message $W_1$ in the BC formed by $X_2$ and the two destinations $(Y,Z)$, however, it creates more interference at user $Z$ that is interested only in message $W_2$. The optimal transmission strategy needs to capture such a tradeoff, and is thus hitherto unknown, however, a few cases denoted as \emph{interference regimes} are to this day fully understood.
 
The first capacity result of this setting is due to Maric \textit{et. al}~\cite{Maric2007} for the ``very strong interference" regime based on an equivalence with the Interference Channel (IFC) with a common message. Later Wu \textit{et. al} in \cite{Wu2007}, and independently Jovicic \textit{et. al} in~\cite{Jovicic2009}, characterized the capacity region of the ``very weak interference" regime. The capacity of the Z-CIFC with a noiseless link was derived by Liu \textit{et al.} in~\cite{LiuZChannel} and the classes of Less-Noisy and More-Capable CIFC were investigated in the works of Vaezi~\cite{VaeziLessNoisy,VaeziMoreCapable}. Lately, Rini~\textit{et. al} proposed in~\cite{Rini2011} a unifying inner bound that is capacity achieving in all regimes in which capacity is known, through a combination of known techniques of binning, rate splitting, and superposition coding. They also suggest a new outer bound that alleviates the computational complexity of existing outer bounds which involve auxiliary random variables. Additionally, based on the proposed inner bound, the capacity region of a new regime, denoted as `` better cognitive decoding" regime, was derived along with the capacity of the semi-deterministic CIFC. 

As for the Gaussian CIFC, the capacity region remains to be fully characterized, yet, some regimes are fully understood: the \emph{weak interference} capacity region was derived in~\cite{Wu2007}, capacity of the \emph{very strong interference} was reported in~\cite{Maric2007} and that of the \emph{primary decodes interference} regime was found in~\cite{RiniGaussian}. The S-CIFC, where interference is experienced only at the primary user's side, was also extensively studied and capacity for the \emph{weak interference} case was characterized under different regimes by Jiang \textit{et al.} in~\cite{JiangMaric}, Vaezi \textit{et al.} in~\cite{VaeziZChannel}, and Rini \textit{et al.} in~\cite{RiniGaussian}. Extensions to multiple users were attempted in the rather exhaustive survey of \cite{Maamari2015} where many messages are to be delivered to many receivers from many transmitters, a subset of which have non-causal knowledge of some messages. 

\begin{figure}[t]
 \centering
 \includegraphics[scale=0.5]{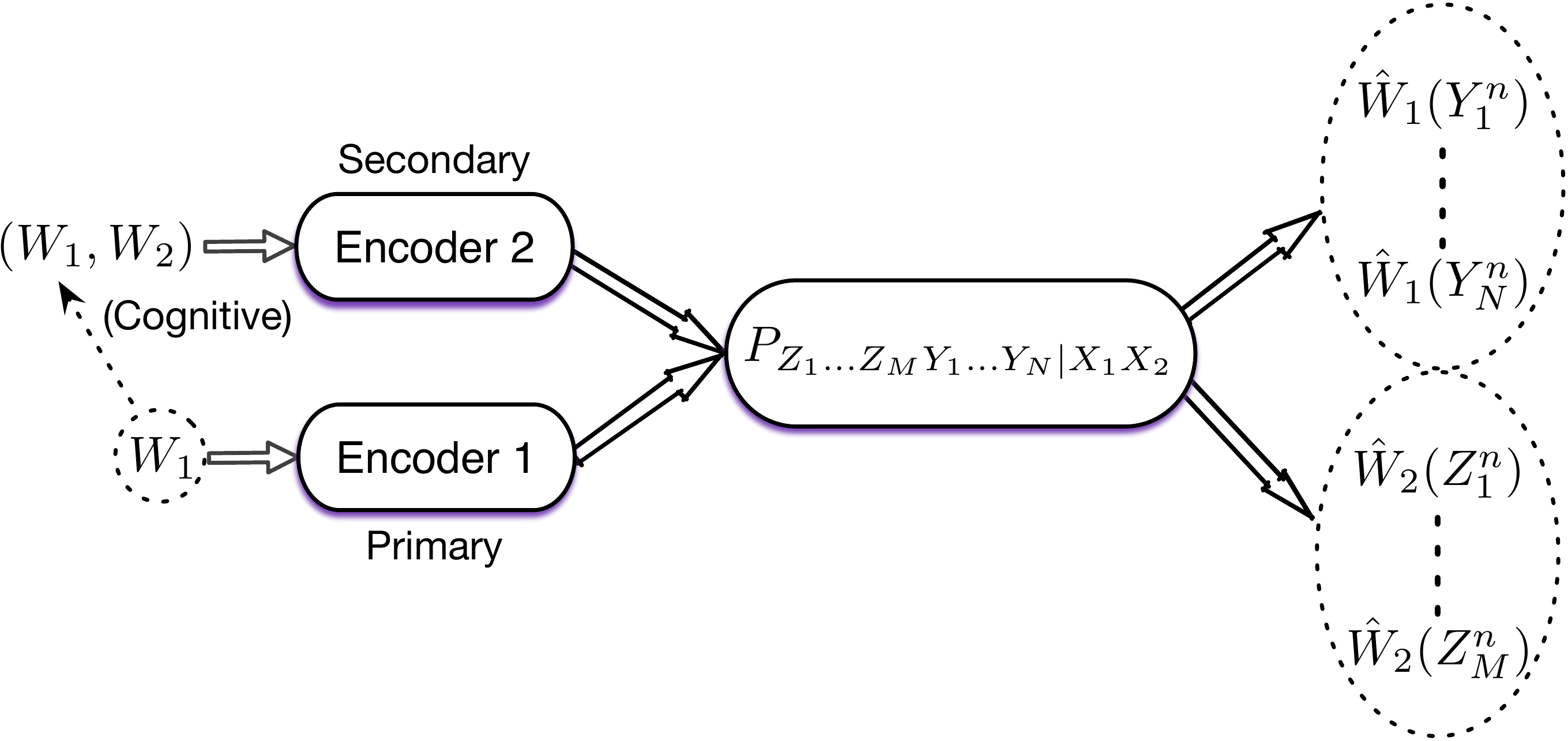} 
 \caption{The Multicast Cognitive Interference Channel.}
 \label{CIFC}
 \end{figure}

In this paper, we investigate another challenging multi-user scenario where \emph{multicasting} transmission over a CIFC  are combined,   as depicted in Fig. \ref{CIFC}. The resulting Multicast CIFC is composed of many primary users interested in a same message $W_1$ and many secondary receivers similarly interested in a same message $W_2$.  
  A possible deployment scenario of this setting consists of several users (spectators) in a football stadium wishing to have access to instantaneous replay of the most important actions on their cell phones while being served by two base stations: a primary one destined to the hosting team, and a secondary one intended  to serve the guest team. 
  
The aim of this work is to infer optimal interference mitigation techniques, considering this multicast setting, through the understanding of how coding schemes should account for multicasting among both primary and secondary receivers. This task is rendered challenging by both the fact that the capacity region for the general CIFC is not known and that multicasting imposes design efficient interference mitigation techniques that could accommodate interference for many users at once. Indeed, even in basic setups of multicast (or compound) Broadcast Channels, i.e. without the helper, optimal schemes remains still unknown. In particular, we investigate two dual classes of Multicast CIFC: the Multi-Primary CIFC where we consider only one secondary receiver and design the optimal schemes to serve a set of $N$ primary receivers, and its dual Multi-Secondary CIFC. The two classes of Multicast CIFC bring distinct challenges to face both at the encoding and at the decoding sides. It is shown that under specific interference regimes, some known optimal schemes without multicasting remain optimal even in the presence of multiple users. Whereas for other interference regimes, the introduction of more evolved encoding and decoding strategies is necessary to derive capacity results.  These strategies consist in the two ideas of ``Interference Decoding" and ``Multiple Description coding"~\cite{BenammarJournalCompound} whose introduction will turn out to yield either optimality result, for the Multi-Primary CIFC, or a significant improvement over standard encoding techniques in the Multi-Secondary CIFC.  
   
 \subsection*{Contributions and outline:}
 We start by deriving in Section \ref{Sec-Definitions} an inner bound to the capacity region of the general Mutlicast CIFC which combines optimal coding techniques of the \emph{Broadcast Channel} (superposition coding and random binning) with rate-splitting for the \emph{Interference Channel}, and which generalizes Marton's inner bound to the Multicast setting.  
Then, in the remainder of the work, we evaluate this rate region in both classes of Multi-Primary and Multi-Secondary CIFC in \emph{very strong interference} and \emph{very weak interference} and a new regime denoted as \emph{mixed very weak/string interference}, as well as the Gaussian counterparts of these regimes. In Section~\ref{Sec-Multi-Primary}, we investigate the Multi-Primary CIFC and derive the following results:  
 \begin{itemize}
 \item The capacity region of the \emph{very strong interference} regime, 
 \item The capacity region of the \emph{very weak interference} regime through a novel outer bound that appears to be necessary to palliate the use of Csisz\'ar \& K\"orner sum-identity,
 \item The capacity region of the \emph{mixed very weak/strong interference} regime while resorting to the idea of \emph{Interference Decoding},
 \item The capacity region of the corresponding Gaussian Multi-Primary CIFC in \emph{very strong interference}, \emph{weak interference}, and \emph{mixed weak/very strong interference} regimes. 
 \end{itemize}
 
 In Section \ref{Sec-Multi-Secondary}, we investigate the Multi-Secondary CIFC and characterize, similarly to the Mutli-Primary CIFC, and derive the following results:  
 \begin{itemize}
 \item The capacity region of the \emph{very strong interference} regime, 
 \item The capacity region of the \emph{very weak interference} regime through a novel outer bound that palliates the use of Csisz\'ar \& K\"orner sum-identity, 
 \item The capacity region of the \emph{mixed very weak/strong interference} regime,  
 \item The capacity region of the corresponding Gaussian Multi-Secondary CIFC in \emph{very strong interference} regime.  
 \end{itemize}

It is worth mention that the Gaussian Multi-Secondary CIFC in the \emph{weak interference} regime appears to be particularly challenging, as shown in the light of the work by Khisti \textit{et al.}~\cite{Khisti2007}. We address this difficulty that arises from multicasting constraint by sending multiple views --intended to different users-- of the same information which resorts to \emph{Multiple Description} coding. Although in this case the  outer and inner bound we derived are not tight, our inner bound strictly outperforms the inner bound where only a common description is used. 

In Section \ref{Sec-Vaezi}, we elaborate on a peculiar fact in the standard (non-multicast) CIFC which is that the \emph{very strong interference} and the \emph{better cognitive decoding} are both included in the \emph{very weak interference} regime as proved by Vaezi in \cite{VaeziWISI}. We show that this claim does not hold in the considered scenario  when we are in the presence of a strict multicast, i.e. $N> 1$ or $M>1$. 
 
\subsection*{Notations and conventions}
 For any sequence~$(x_i)_{i\in\mathbb{N}_+}$, notation $x_k^n$ stands for the collection $(x_k,x_{k+1},\dots, x_n)$. $x_1^n$ is simply denoted by $x^n$. Entropy is denoted by $H(\cdot)$, and mutual information by $I(\cdot;\cdot)$. $\mathds{E}$ resp. $\mathds{P}$ denote the expectation w.r.t. the generic probability distributed (PD) while the notation $P$ is specific to the probability of a random variable (rv). Let $X$ and $Y$ be random variables, we denote by $P_{XY}$ (resp. $P_{Y|X}$, and $P_X$) the joint probability distribution of $(X,Y)$ (resp. conditional distribution of $Y$ given $X$, and marginal distribution of $X$). $\|\cX\|$ stands for the cardinality of the set $\cX$. We denote typical and conditional typical sets by $\typ{X}$ and $\typ{Y|x^n}$, respectively (see Appendix~\ref{sec:typical} for details). Let $X$, $Y$ and $Z$ be three random variables on some alphabets with probability distribution~$p$. If $p(x|yz)=p(x|y)$ for each $x,y,z$, then they form a Markov chain, which is denoted by $X\mkv Y\mkv Z$. 
 %
 %
\section{Problem Definition and Inner bound}\label{Sec-Definitions}
The discrete memoryless N-Multicast CIFC can be represented by the conditional pmf:  
\begin{equation}
 P_{Z_1^n \cdots Z_M^n Y_1^n \cdots Y_N^n |X_1^n X_2^n} = \prod_{i= 1}^n P_{Z_{1,i} \cdots Z_{M,i} Y_{1,i} \cdots Y_{N,i}| X_{1,i} X_{2,i}} \ . 
\end{equation}

An ${(M_{1n}, M_{2n}, n)}$-code for this channel consists of: two sets of messages, $\cM_1\equiv \{1,\dots,M_{1n}\}$ and $\cM_2\equiv  \{1,\dots,M_{2n}\}$, two encoding functions, and $(N+M)$-decoding functions. The encoding function at source $1$ (the primary source) assigns an n-sequence $x^n_1 (W_1) $ to each message $W_1$ while the encoding function at source $2$ assigns an $n$-sequence $x^n_2 ( W_1, W_2)$ to each pair of messages $(W_1, W_2) \in  \cM_1 \times \cM_2 $. The secondary decoders indexed by $k \in [1:M]$ assign to each received sequence $Z_k^n$ an estimate message $  \hat{W}_{2,k} $ while the primary decoders indexed by $j \in[1:N]$ each resort to a decoding function that assigns to each received sequence $Y_j^n$, an estimate message $\hat{W}_{1,j}$.	
	
	The probability of error is given by: 
\begin{IEEEeqnarray}{rCl}	  
 P_e^{(n)}  &\equiv & \mathds{P} \biggl(  \bigcup_{j \in [1:N]} \bigl\{ \hat{W}_{1,j} \neq W_1 \bigr\} \text{ or } \bigcup_{k \in [1:M]}   \bigl\{\hat{W}_{2,k} \neq W_2 \bigr\}\biggr)\ . 
\end{IEEEeqnarray}
	  A rate pair $(R_1, R_2)$ is said to be achievable if there exists an ${(M_{1n}, M_{2n}, n)}$-code satisfying: 
\begin{IEEEeqnarray}{rCl}
	 \liminf\limits_{n \rightarrow \infty} \frac{1}{n} \log_2 M_{ln} &\geq & R_l\ , \,\,\,\textrm{ $l=\{1,2\}$}\,,\\
	 \limsup\limits_{n \rightarrow \infty}  P_e^{(n)}  &=& 0\ . 
\end{IEEEeqnarray}
The convex closure over all achievable rate pairs $(R_1,R_2)$ defines the capacity region. 
   
\subsection*{Inner bound to the capacity region of the Multicast  CIFC}

In this section, we derive an inner bound to the capacity region of the CIFC, that generalizes Marton's inner bound for the BC in the presence of a helper, as shown in Fig. \ref{CIC}. 
  
 \begin{theorem}[Inner bound]\label{theorem-inner-bound}
 An inner bound to the capacity region of the CIFC consists in all rate pairs $(R_1,R_2)$ satisfying:
 \begin{IEEEeqnarray}{rCl}\label{CDcoding}
 \IEEEyesnumber\IEEEyessubnumber*
 R_1 &\leq& \min_{j\in[1:N]} I(Q_1X_1Q U; Y_j)\\ 
 R_2 &\leq&  \min_{k\in[1:M]}  I(Q V; Z_k|Q_1) - I(Q V; X_1|Q_1) \\
 R_2 &\leq& \min_{j\in[1:N]}I(X_1 U; Y_j | Q_1 Q) + \min_{k\in[1:M]} I( Q V; Z_k|Q_1) - I(V; X_1 U | Q_1Q ) \\ 
 R_2 &\leq& \min_{j\in[1:N]}I(X_1 QU; Y_j | Q_1 ) + \min_{k\in[1:M]} I( V; Z_k|Q_1Q ) - I(V; X_1 U | Q_1Q ) \\ 
 R_2 &\leq& \min_{j\in[1:N]}I(X_1 QU; Y_j | Q_1 ) + \min_{k\in[1:M]} I( Q V; Z_k|Q_1) \nonumber\\
 &-& I(V; X_1 U | Q_1Q ) - I(Q;X_1|Q_1) \qquad \ \\ 
 R_1 + R_2 &\leq& \min_{j\in[1:N]}I(X_1 U; Y_j | Q_1 Q) + \min_{k\in[1:M]} I(Q_1 Q V; Z_k) - I(V; X_1 U | Q_1Q ) \\
 R_1 + R_2 &\leq& \min_{j\in[1:N]}I(Q_1X_1 QU; Y_j) + \min_{k\in[1:M]} I(V; Z_k|Q_1Q) - I(V; X_1 U | Q_1Q ) \\
 R_1 + R_2 &\leq& \min_{j\in[1:N]}I(Q_1X_1 QU; Y_j) + \min_{k\in[1:M]} I( QV; Z_k|Q_1) \nonumber\\
 &-& I(V; X_1 U |Q_1 Q ) - I(Q;X_1|Q_1) \qquad  \\
 R_1 + R_2 &\leq& \min_{j\in[1:N]}I(X_1 QU; Y_j |Q_1) +\min_{k\in[1:M]} I(Q_1 QV; Z_k ) \nonumber\\
 &-& I(V; X_1 U |Q_1 Q ) - I(Q;X_1|Q_1)\qquad  \\
 R_1 + 2 R_2 &\leq& \min_{j\in[1:N]}I(X_1 QU; Y_j|Q_1) + \min_{k\in[1:M]} I( Q_1QV; Z_k) \nonumber\\
 &+&  \min_{k\in[1:M]} I(V;Z_k|Q_1Q)- I(V; X_1 U |Q_1 Q )  - I(Q;X_1|Q_1) \\
 R_1 + 2 R_2 &\leq& \min_{j\in[1:N]}I(Q_1X_1 QU; Y_j) + \min_{k\in[1:M]} I( QV; Z_k|Q_1) \nonumber\\
 &+& \min_{k\in[1:M]} I(V;Z_k |Q_1Q)- I(V; X_1 U |Q_1 Q )  - I(Q;X_1|Q_1)  \ , 
 \end{IEEEeqnarray} for some joint PD $P_{Q_1X_1QUVX_2}$ satisfying $ (Q_1 Q U V) \mkv (X_1,X_2) \mkv (Y_1,\cdots , Y_N, Z_1, \cdots, Z_M)$.
\end{theorem}
 
\begin{remark}
In the absence of the helper, i.e.,  when $X_1 = Q_1 = \emptyset$, the inner bound collapses to Marton's inner bound in the multicast setting with the common auxiliary rv $Q$ and the two private ones $U$ and $V$. On the other hand, the variables $X_1$ and $Q_1$ account for rate-splitting at the primary source. The rate splitting at the secondary source is already contained in Marton's coding~\cite{MartonInnerBound79} with the common auxiliary $Q$. Thus, this inner bound combines both optimal coding schemes for the Broadcast Channel~\cite{MartonInnerBound79} and Interference Channel~\cite{1056307}. 
 \end{remark} 
 \begin{IEEEproof}The proof is relegated to Appendix~\ref{ProofOfAchievabilityMulticast}.\end{IEEEproof}

In the sequel, we investigate two distinct classes of Multicast-CIFC. The Multi-Primary CIFC where we assume the existence of only one secondary receiver $M=1$, which will allow us to investigate the optimal interference mitigation techniques for primary receivers. Then,  we investigate as well the dual class referred  to as  Multi-Secondary CIFC where we consider only one primary receiver $N=1$ and characterize the optimal interference mitigation technique within the set of secondary users.  
  
 %
 %
\section{Capacity results for the Multi-Primary CIFC}\label{Sec-Multi-Primary}
 
\begin{figure}[h!]
\centering
\includegraphics[scale=.5]{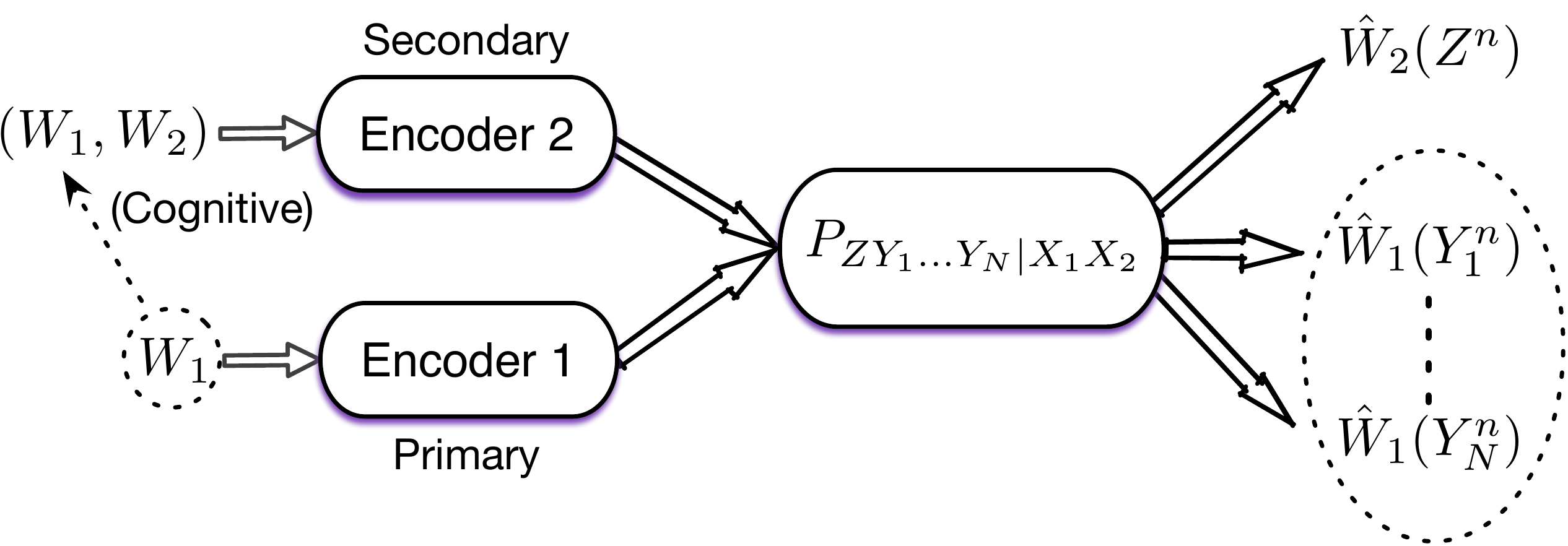}
\caption{The Multi-Primary Cognitive Interference Channel.}
\label{MultiPrimaryCIFC}
\end{figure}

We refer to this class of Multicast CIFC as the Multi-Primary CIFC since it consists of only one secondary receiver, $Z$ as denoted in the sequel, and $N$ primary receivers as shown in Fig.~\ref{MultiPrimaryCIFC}. 

Our aim in this section is to characterize the capacity region of this class of channels under distinct interference regimes resorting to the inner bound stated in Theorem\ref{theorem-inner-bound} and, in a specific regime, a more evolved decoding scheme --referred to as Interference Decoding-- when \eqref{CDcoding} would be limited by the decoding behavior of the users. 
  \subsection{Capacity region of the Multi-Primary CIFC in the Very Strong Interference regime}
  In this part of the work, we derive the capacity region of the Multi-Primary CIFC setting in the \emph{very strong interference} regime.  
  To this end, consider the multicast CIFC described in Fig~\ref{MultiPrimaryCIFC}. 
  
  \begin{assumption}[Strong interference]
The \emph{strong interference} condition is defined as: 
  \begin{equation} \label{StrongIFcondition}
 \forall P_{X_1X_2} \ ,\qquad  I(X_2;Z|X_1) \leq \displaystyle\min_{j\in [1:N]}   I(X_2;Y_j|X_1)  \ .
  \end{equation}
 \end{assumption}
   
  \begin{assumption}[Very strong interference]
    The \emph{very strong interference} condition is defined as: 
  \begin{equation}\label{VeryStrongIFcondition}
   \forall  P_{X_1X_2} \ ,\qquad  \displaystyle\min_{j\in [1:N]}    I(X_1X_2;Y_j)  \leq  I( X_1X_2;Z)  \ .
  \end{equation}
   \end{assumption}

  \begin{theorem}[Very strong interference]
  The capacity region of the Multi-Primary CIFC satisfying assumptions \eqref{StrongIFcondition} and \eqref{VeryStrongIFcondition} is given by the set of rate pairs $(R_1,R_2)$ that verify: 
  \begin{IEEEeqnarray}{rCl} 
  R_2 &\leq& I(X_2;Z|X_1)  \ , \\ 
  R_1 + R_2 &\leq& \displaystyle\min_{j\in [1:N]}   I(X_1X_2;Y_j)  \ ,
  \end{IEEEeqnarray}  
for some PD $P_{X_1X_2}$. 
  \end{theorem}
 \begin{IEEEproof}
  The proof of achievability follows from rate region \eqref{CDcoding} letting $Q_1 = X_1$ and $Q= U= V = X_2$. This implies that all users $Y_j$ and $Z$ decode both signals $(X_1,X_2)$.  
 
  The obtained rate region is of the form: 
    \begin{IEEEeqnarray}{rCl}
     \IEEEyesnumber\IEEEyessubnumber*
   R_1 &\leq& \displaystyle\min_{j\in [1:N]}  I(X_1X_2;Y_j)  \ , \label{Cstr1}  \\
   R_2 &\leq& I(X_2;Z|X_1)  \ , \\ 
   R_2 &\leq& \displaystyle\min_{j\in [1:N]}  I(X_2;Y_j|X_1) \ , \label{Cstr2}  \\
   R_1 + R_2 &\leq& I(X_1X_2;Z)  \ , \label{Cstr3}  \\
   R_1 + R_2 &\leq& \displaystyle\min_{j\in [1:N]}  I(X_1X_2;Y_j)\label{Cstr4}  \ .
  \end{IEEEeqnarray}
 One can notice that \eqref{Cstr1} is redundant due to the sum-rate \eqref{Cstr4} while \eqref{Cstr2} is redundant due the strong interference assumption~\eqref{StrongIFcondition}, and expression~\eqref{Cstr3} is redundant by the \emph{very strong interference} condition~\eqref{VeryStrongIFcondition}. As for the converse, it follows from the \emph{strong interference} condition  \eqref{StrongIFcondition} alone and is given in Appendix \ref{App-VSI-MultiPrimary}. 
  \end{IEEEproof}  
 
   \subsection{Capacity region of the Multi-Primary CIFC in the Very Weak Interference regime}
   
We now give the capacity region of the under the \emph{very weak interference regime}.  To this end, se state the next assumption. 

\begin{assumption}[Weak interference]
Let $\mathcal{P}_w$ be the set of all joint PDs $P_{UX_1X_2}$ which satisfy the following Markov chain: 
  \begin{equation}
   U \mkv (X_1,X_2) \mkv (Y_1,\cdots, Y_N, Z_1 , \cdots, Z_M) \ . 
  \end{equation}
The \emph{weak interference} regime is defined as: 
   \begin{equation}\label{WeakInterferenceCondition}
 \forall j\in [1:N]\ , \quad  \forall P_{UX_1X_2} \in \mathcal{P}_w \ ,\quad I(U; Y_j | X_1 )\leq I(U;Z | X_1)  \ . 
\end{equation}
 \end{assumption}
 
 \begin{assumption}[Very weak interference]
The \emph{very weak interference} regime is to verify~\eqref{WeakInterferenceCondition} while to further satisfy:
     \begin{equation}\label{VeryWeakInterferenceCondition}
 \forall j\in [1:N] \ ,\quad  \forall P_{UX_1X_2} \in \mathcal{P}_w \ , \quad I(UX_1 ; Y_j   ) \leq I(U X_1;Z  )  \ . 
\end{equation}
\end{assumption}

 The intuitive extension of the result of the standard CIFC (non-multicast CIFC with $N=1$)  to an arbitrary number of users $N$ will prove to be optimal for this regime as well.  Yet, the proof of the resulting capacity region presents a major difficulty in the converse part. Indeed, the multicast nature of the transmission prevents from resorting to Csisz\'ar \& K\"orner's identity that was originally used in \cite{Jovicic2009,Wu2007}. To this end, an alternative proof of converse is required herein.   

  \begin{theorem}[Very weak interference]
  The capacity region of the Multi-primary CIFC in \emph{very weak interference} regime is given by the set of rate pairs satisfying: 
  \begin{IEEEeqnarray}{rCl}
    R_1 &\leq& \displaystyle\min_{j\in [1:N]}   I(X_1U;Y_j)   \ , \\ 
    R_2 &\leq& I(X_2;Z|X_1U) \ ,  
  \end{IEEEeqnarray} 
for some arbitrarily PD $P_{UX_1X_2}$ satisfying $ U\mkv (X_1,X_2) \mkv (Y_1,\cdots , Y_N, Z)$. 
  \end{theorem}
  \begin{IEEEproof}
  The achievability follows from the inner bound in~\eqref{CDcoding} by letting $Q = U$, $Q_1 = (X_1,U)$, and $V = X_2$ which amounts to enabling decoders $Y_j$ to decode only their useful signal while user $Z$ decodes both signals $(X_1,X_2)$. 
We end up with the inner bound: 
\begin{IEEEeqnarray}{rCl}
R_1 &\leq& \displaystyle\min_{j \in[1:N]} I(X_1 U;  Y_j)  \ , \\
R_2 &\leq&  I(X_2;Z |  X_1U) \ ,  \\ 
R_1 + R_2 &\leq& I( X_1 X_2;Z) \ . 
\end{IEEEeqnarray}  
Using the \emph{very weak interference} condition \eqref{VeryWeakInterferenceCondition}, the sum-rate bound is redundant. Hence, the achievability is proved. As for the converse, the proof is more evolved and is presented in Appendix~\ref{App-VWI-MultiPrimary}. The proof is valid under the loose condition of \emph{weak interference} assumption alone~\eqref{WeakInterferenceCondition} and originates from noticing that this  condition implies a ``conditional less-noisiness" \cite{NairLessNoisy} order between $Y_j$ and $Z$ conditioned on $X_1$. 
  \end{IEEEproof}
  
     In both previously investigated regimes of \emph{very weak} and \emph{very strong} interference, the inner bound \eqref{CDcoding} proves to be optimal and an extension of the intuitive coding schemes to an arbitrary number of users turns out to be optimal, though it might have required to write a new outer bound to prove the optimality. In the sequel, we investigate a setting that did not arise in the non-multicast CIFC where $N=1$ and that will turn out to require a more evolved coding scheme that can be encompassed with the inner bound~\eqref{CDcoding}. 
 \subsection{Capacity region of the Multi-Primary CIFC in the Mixed Very Weak/Strong Interference regime}

We consider a Multi-Primary CIFC where we can partition the multicast set of users $Y_j$ , $j \in [1:N]$, into two subsets. A subset $\mathcal{W}$ where all users are in \emph{weak interference}, and a subset $\mathcal{S}$ where users are in the \emph{strong interference} regime, however, \emph{very weak or very strong interference} can be verified in either of the sets, without being imposed on both sets. 

\begin{assumption}[Mixed very weak/strong interference]
The \emph{mixed very weak/strong interference} regime is defined by the following conditions: 
 \begin{IEEEeqnarray}{rCl}
   \forall j \in \mathcal{W} \quad , \quad I(U; Y_j |X_1) &\leq& I(U; Z |X_1) \ , \label{Mixed1}\\
   \forall j \in \mathcal{S} \quad , \quad I(X_2; Z|X_1) &\leq& I(X_2; Y_j |X_1) \ ,  \\
    \min_{j\in \mathcal{S}} I(X_1 X_2; Y_j) \leq I(X_1 X_2; Z) \  &\text{ or }&  \  \min_{j\in \mathcal{W}} I(UX_1; Y_j) \leq I(U X_1 ; Z)\label{Mixed4} \ , 
   \end{IEEEeqnarray}
for all $P_{UX_1X_2}$ that satisfied the Markov chain:  
   \begin{equation}
   U \mkv (X_1,X_2) \mkv (Y_1,\cdots, Y_N, Z_1 , \cdots, Z_M) \ . 
  \end{equation}
  \end{assumption}
  
  Before enunciating the main result of this section, the following remark is crucial. The set of users $Y_j$ that experience \emph{very strong interference} would require decoding the interfering codeword $X_2$, whilst those that experience \emph{very weak interference} are required to decode only their intended useful codewords $U$ and $X_1$. User $Z$ is required to decode all codewords in both interference regimes. As it is enunciated, the inner bound~\eqref{CDcoding} fails in encompassing this differentiated behaviour of users $Y_j$ since it relies on a similar decoding strategy for all users $Y_j$. Thus, the need for a more evolved decoding strategy arises, which justifies the introduction of the idea of ``Interference Decoding".  
  
   This decoding strategy was first investigated by the authors for a \emph{Compound Broadcast Channel} \cite{BenammarJournalCompound} which can be regarded as a multicast setting where the many users of the multicast are designated by the many possible channel instances of the Broadcast setting. \emph{Interference Decoding} allows each user in the multicast group $Y_j$ to decode/ or not the interfering codeword $X_2$, and thus, provides for a differentiated decoding at all users. Resorting to this idea allows to state the following result. 
   
   \begin{theorem}[Mixed very weak/strong interference]
   The capacity region of the Multi-primary CIFC satisfying conditions \eqref{Mixed1}--\eqref{Mixed4}, is given by the set of rate pairs satisfying: 
   \begin{IEEEeqnarray}{rCl}
   R_1 &\leq& \displaystyle \min_{j \in \mathcal{W}} I(U X_1;Y_j) \ ,  \\
   R_2 &\leq& I(X_2; Z | UX_1) \ ,  \\
   R_1 + R_2 & \leq &  \displaystyle\min_{j \in \mathcal{S}} I(X_1 X_2;Y_j) \ ,
   \end{IEEEeqnarray} 
   for some joint input PD $P_{U X_1X_2}$ such that $ U\mkv (X_1,X_2) \mkv (Y_1,\cdots , Y_N, Z)$.
   \end{theorem}
   
\begin{IEEEproof}
     The converse proof follows in the exact same manner as the converse proof of both weak and strong interference cases. As for the achievability part, it is more involved and requires introducing the idea of \emph{Interference Decoding}. The decoders $Y_j$ with $j \in \mathcal{W}$ will choose to decode only the useful signal $U$ and $X_1$ while the users that are in strong interference, i.e  $Y_j$ with $j \in \mathcal{S}$, will decode all signals transmitted by source $2$ and source $1$, that is: $(U,X_1,X_2)$. User $Z$ is allowed to decode all codewords as well.   Proof details can be found in Appendix \ref{App-Mixed-MultiPrimary}.     
     \end{IEEEproof}
   
  \subsection{Capacity results for the Gaussian Multi-Primary CIFC}
\begin{figure}[h!]
 \centering
 \includegraphics[scale=0.6]{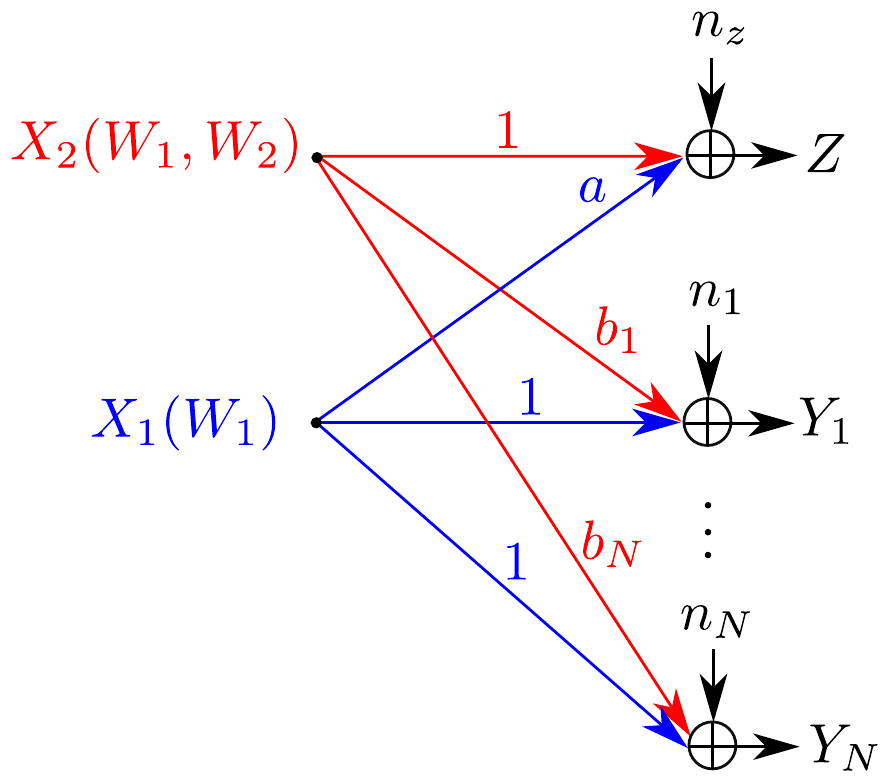} 
 \caption{The Gaussian Multi-Primary Cognitive Interference Channel.}
 \label{GaussianCIFC} 
 \end{figure}

  Consider the following Gaussian Multicast CIFC model as shown in Fig \ref{GaussianCIFC}: 
  \begin{eqnarray}
 \forall j\in [1:N] \ ,\quad  Y_j &=& b_j X_2 + X_1 + n_j  \ , \\   
  Z &=& X_2 + a X_1 + n_z \ , 
  \end{eqnarray}
  where $b_j$, $j \in [1:N]$, and $a$ are real numbers, and where $(n_1, \cdots, n_N)$, and $n_z$ are additive white Gaussian noise components with powers $N_1 = \cdots = N_N = N_z = 1$. 
 
In this section, we derive the capacity region for several different regimes of the Gaussian Multi-Primary CIFC. 
  
  \begin{theorem}[Gaussian Multi-Primary CIFC] 
  \label{TheoremGaussianMultiPrimaryCIFC}
  The capacity region of the Gaussian Multi-Primary CIFC is characterized in the following regimes: 
  \begin{itemize}
  \item[1)]   The capacity region of the \emph{very strong interference} regime  is given by the set of rate pairs that satisfy:
  \begin{IEEEeqnarray}{rCl}
   \forall j \in [1:N] \ &,& \ |b_j| \geq 1  \ , \\
  \forall \rho \in [-1:1]  \ &,& \  \displaystyle\min_{j\in [1:N]} \biggl\{ (1-a^2)P_1 +(b_j^2-1)P_2 + 2\rho( b_j-a) \sqrt{P_1P_2} \biggr\} \leq  0  \ ,  
   \end{IEEEeqnarray}  
  consists in the set of rate pairs satisfying: 
\begin{IEEEeqnarray}{rCl}
R_2 &\leq&  \dfrac{1}{2} \log_2 \left( 1 +  (1 - \rho^2) P_2 \right)  \ , \\
R_1 + R_2 &\leq& \dfrac{1}{2} \displaystyle\min_{j\in[1:N]}  \log_2 \left(  1 + b_j^2 P_2 + P_1 + 2 b_j\rho \sqrt{ P_1 P_2} \right) \ , 
\end{IEEEeqnarray}   
for some $\rho \in [-1:1]$. 
  
  \item[2)]  The capacity region of the \emph{weak interference regime}  is given by the set of rate pairs that satisfy:
  \begin{equation}
   \forall j \in [1:N] \ , \  |b_j| \leq 1  \ ,  
  \end{equation}
  is given by the set of rate pairs satisfying: 
\begin{IEEEeqnarray}{rCl}
R_1 &\leq&  \dfrac{1}{2} \ \displaystyle\max_{ \rho \in[-1:1]} \ \displaystyle\min_{j\in[1:N]} \ \log_2 \left( \dfrac{1 + b_j^2 P_2 + P_1 + 2 b_j\rho \sqrt{ (1- \eta )P_1 P_2}}{ 1 + b_j^2 \eta P_2} \right)  \ , \\
R_2 &\leq&   \dfrac{1}{2} \log_2 \left( 1 + \eta P_2  \right) \ , 
\end{IEEEeqnarray}  
for some $\eta \in [0:1]$.
  \item[3)] The capacity region of the \emph{mixed weak/very strong interference} regime, where  $[1:N] = \mathcal{S}\cup \mathcal{W}$ and 
  \begin{IEEEeqnarray}{rCl}
   \forall j \in  \mathcal{W} \ &,& \ |b_j| \leq 1  \ , \\
   \forall j \in  \mathcal{S} \ &,& \ |b_j| \geq 1  \ , \\
  \forall \rho \in [-1:1]  \ &,& \  \displaystyle\min_{j\in  \mathcal{S}} \biggl\{ (1-a^2)P_1 +(b_j^2-1)P_2 + 2\rho( b_j-a) \sqrt{P_1P_2} \biggr\} \leq  0  \ , 
   \end{IEEEeqnarray}
 is given by the set of rate pairs that satisfy: 
  \begin{IEEEeqnarray}{rCl}\label{CapacityMixed}
R_1 &\leq&  \dfrac{1}{2} \ \displaystyle\min_{j \in  \mathcal{W}} \ \log_2 \left( \dfrac{1 + b_j^2 P_2 + P_1 + 2 b_j\rho \sqrt{ (1- \eta )P_1 P_2}}{ 1 + b_j^2 \eta P_2} \right)  \ , \\
R_2 &\leq& \dfrac{1}{2} \log_2 \left( 1 + \eta P_2  \right) \ , \\
R_1  + R_2 &\leq& \dfrac{1}{2} \ \displaystyle\min_{j \in  \mathcal{S}} \ \log_2 \left(  1 + b_j^2 P_2 + P_1 + 2 b_j\rho \sqrt{ (1- \eta )P_1 P_2} \right) \ ,  
\end{IEEEeqnarray} 
for some $\eta \in [0:1]$ and $\rho \in[-1:1]$.
  \end{itemize}
  \end{theorem}
  
\begin{IEEEproof}
A detailed proof of this theorem is given in Appendix \ref{App-Gaussian-MultiPrimary}. However, hereafter the outline of proof for every interference regime.
\begin{enumerate} 
\item \emph{Very strong interference regime}. The achievability scheme consists in letting the two Gaussian codewords $(X_1,X_2)$ to be arbitrarily correlated, defining thus: $ \mathbb{E}[X_1 X_2] = \rho \sqrt{P_1P_2}$. The proof of converse can be deduced from Rini \textit{et al.}'s work \cite{RiniGaussian} by showing that their suggested outer bound can be generalized to an arbitrary number of primary users. 
 
\item \emph{Weak interference regime}. Note here first that we no longer consider \emph{very weak interference} but we rather investigate the looser condition of \emph{weak interference}. The reason being that, in the Gaussian case, we can resort to Dirty-Paper Coding to eliminate the interference $(U, X_1)$ at decoder $Z$, alleviating thus the decoding constraint at user $Z$, i.e.,  
 \begin{equation}
  I(UX_1;Y_j)  \leq  I(UX_1; Z)\ . 
 \end{equation}
 The achievability part is  briefly described  as follows: Split the power $P_2$ into two parts  $\eta P_2$ and $(1- \eta) P_2$. Generate a Gaussian ARV $U$ with variance  $(1- \eta) P_2$ and let it arbitrarily correlated to $X_1$ through the correlation coefficient $\rho$. Generate a Gaussian ARV $V$ that dirty paper codes jointly the interference component $X_u + a X_1$. The converse part follows from \emph{Entropy Power Inequality} along with some analytic manipulations. 
 
\item \emph{Mixed weak/very strong interference regime}. The achievability relies on the idea of Interference Decoding along with a strategic combination of both coding schemes for weak and very strong interference regimes. As for the converse, it follows similarly from combining entropy power inequality and entropy maximization techniques. 
 \end{enumerate}
\end{IEEEproof}
 
In the sequel we give a simplified expression of the capacity regions enunciated above for some setting where the interference components seen at the primary decoders are \emph{coherent}, i.e. all $b_j$'s are of the same sign, as opposed to the non-coherent interference case where $b_j$'s might be of distinct signs.
  
\begin{corollary} 
1) Provided that all gains $b_j$'s share the same sign, i.e.,  coherent weak interference, the capacity region in weak interference regime is given by the set of rate pairs satisfying: 
\begin{equation}
\mathcal{C}_{WI}(\eta)\ : \ \left\{\begin{array}{rcl}
R_1 &\leq&  \dfrac{1}{2} \  \displaystyle \min_{j\in[1:N]} \ \log_2 \left( \dfrac{1 + b_j^2 P_2 + P_1 + 2 | b_j| \sqrt{ (1 -\eta)P_1 P_2}}{ 1 + b_j^2 \eta P_2} \right)  \ , \\
R_2 &\leq&   \dfrac{1}{2} \log_2 \left( 1 + \eta P_2  \right) \ .
\end{array}\right.
\end{equation}  
Moreover, the capacity region consists of the intersection of all capacity regions of the CIFCs $(Z,Y_j)$.
 
2) Provided that all gains $b_j$'s share the same sign, i.e., coherent strong interference,  the capacity region is given by the set of rate pairs satisfying: 
\begin{equation}
\mathcal{C}_{SI}(\eta)\ : \ \left\{\begin{array}{rcl}
R_2 &\leq&   \dfrac{1}{2} \log_2 \left( 1 + \eta P_2  \right) \ ,  \\
R_1 +R_2&\leq&  \dfrac{1}{2} \  \log_2 \left(  1 + b_*^2 P_2 + P_1 + 2 | b_*| \sqrt{ P_1 P_2}  \right)  \ , 
\end{array}\right.
\end{equation}  
where 
\begin{equation}
 | b_*|  \equiv   \displaystyle \min_{j\in[1:N]}  |b_j|  \ . 
\end{equation} 
Moreover, the capacity region consists in the intersection of all capacity regions of the CIFCs $(Z,Y_j)$. 

\begin{figure}[ht]
\centering
\includegraphics[scale=.7]{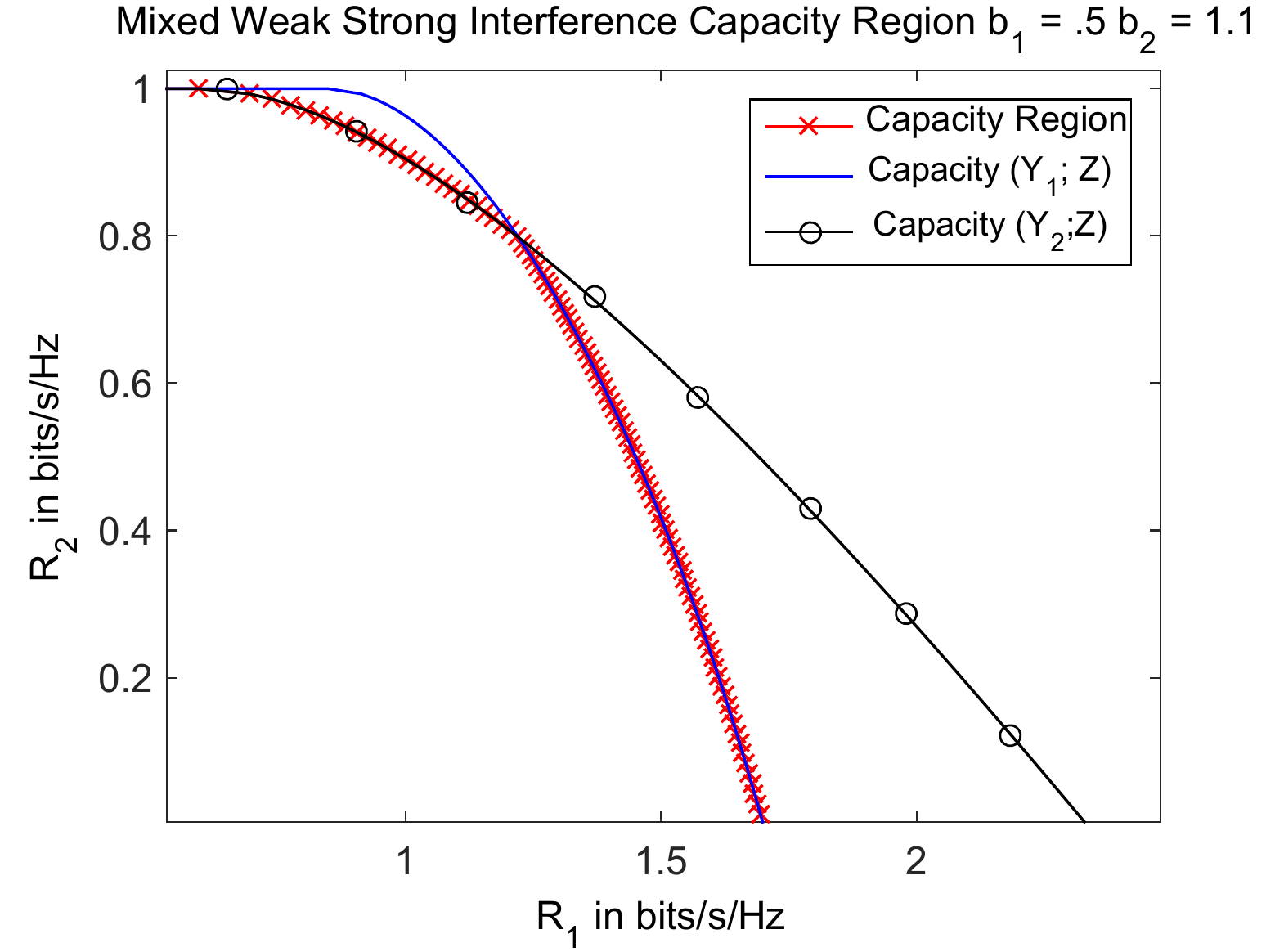}
\caption{The Coherent Multi-Primary Cognitive Interference Channel.}
\label{Coherent}
\end{figure}
\begin{figure}[ht]
\centering
\includegraphics[scale=.7]{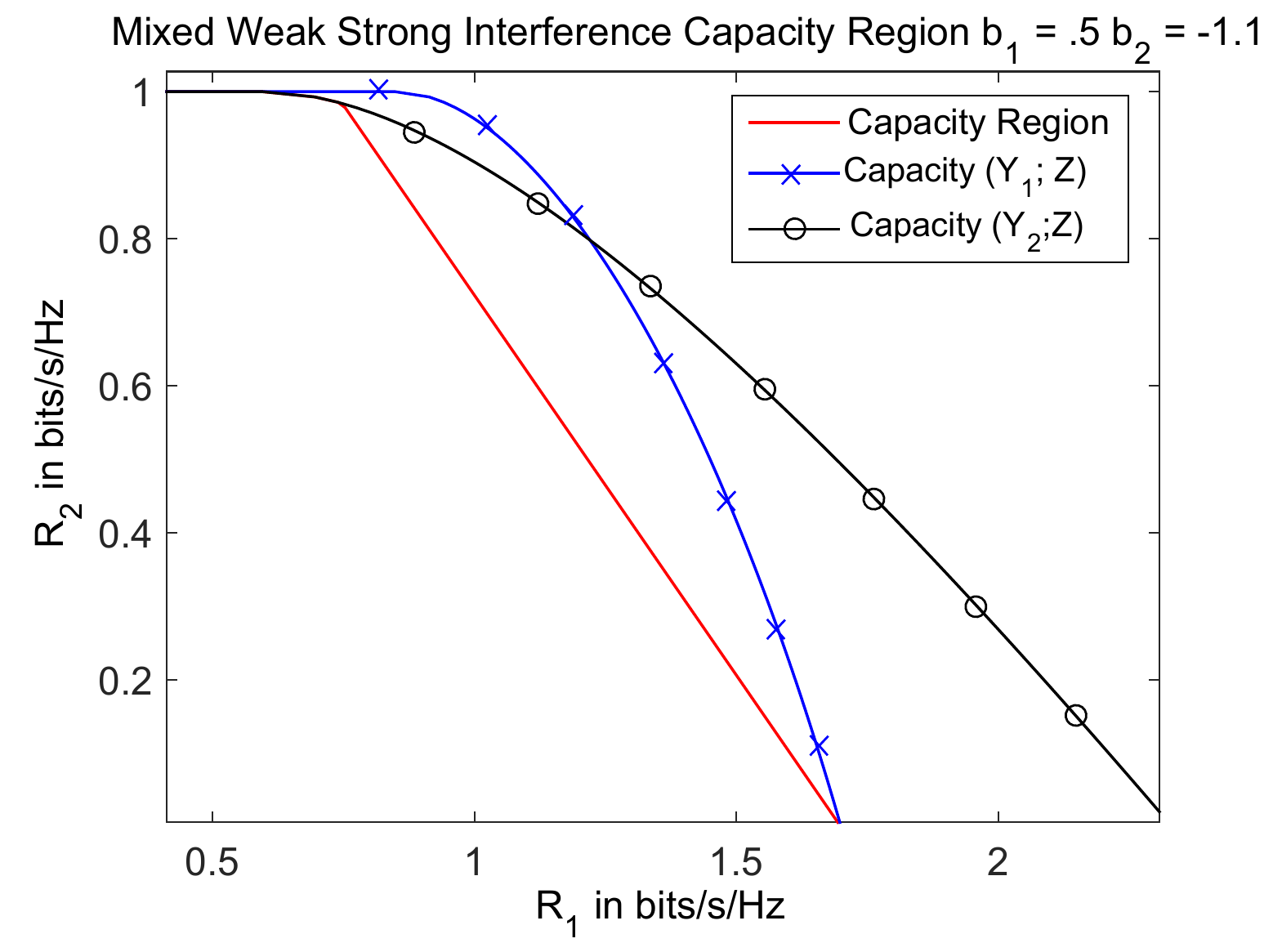}
\caption{The Non-Coherent Multi-Primary Cognitive Interference Channel.}\label{NonCoherent}
\end{figure} 

3 ) Provided that all gains $b_j$'s share the same sign,  i.e., coherent mixed interference,  the capacity region is given by the set of rate pairs satisfying: 
  \begin{equation} 
\mathcal{C}_{MI}(\eta)\ : \left\{\begin{array}{rcl}
R_1 &\leq&  \dfrac{1}{2} \ \displaystyle\min_{j \in  \mathcal{W}} \ \log_2 \left( \dfrac{1 + b_j^2 P_2 + P_1 + 2 |b_j| \sqrt{ (1- \eta )P_1 P_2}}{ 1 + b_j^2 \eta P_2} \right)  \ , \\
R_2 &\leq& \dfrac{1}{2} \log_2 \left( 1 + \eta P_2  \right) \ , \\
R_1  + R_2 &\leq& \dfrac{1}{2} \ \displaystyle\min_{j \in  \mathcal{S}} \ \log_2 \left(  1 + b_j^2 P_2 + P_1 + 2| b_j|  \sqrt{ (1- \eta )P_1 P_2} \right)  \ .
\end{array}\right.
\end{equation}  
In addition, the capacity region is the intersection of the single capacity regions. 
\end{corollary}

\begin{IEEEproof}
The proof of this corollary is given in Appendix~\ref{App-Gaussian-MultiPrimary}. 
\end{IEEEproof} 

To illustrate the results of this corollary, we plot the capacity region for two settings, in both \emph{coherent} and \emph{non-coherent} interference regimes as shown in Fig.~\ref{Coherent} and Fig.~\ref{NonCoherent}.

In this section, we have investigated the Multi-Primary CIFC which consists of one secondary user $Z$ and $N$ primary users $Y_j$, $j \in [1:N]$. The aim was to characterize the capacity region in different interference regimes and infer on the optimal required  interference mitigation techniques. The need for a more evolved decoding technique arises in the \emph{mixed interference} case, as opposed to the \emph{very weak} and \emph{very strong interference} regimes where the intuitive extension of the optimal non-multicast scheme turned out to be optimal for the multicast setting as well. In the sequel, we investigate the Multi-Secondary CIFC and characterize in a similar manner the optimal interference mitigation techniques in distinct interference regimes.

 %
 %
\section{Capacity results for the Multi-Secondary CIFC}\label{Sec-Multi-Secondary}
 
\begin{figure}[h!]
\centering
\includegraphics[scale=.5]{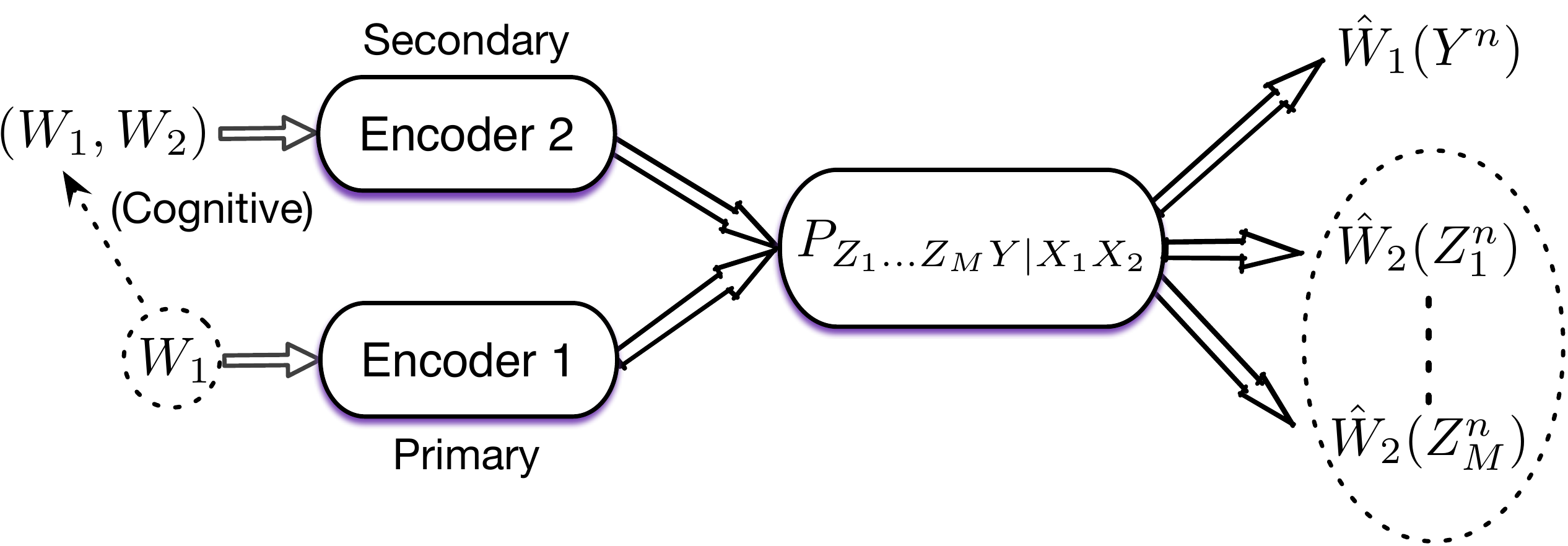}
\caption{The Multi-Secondary Cognitive Interference Channel.}
\label{MultiSecondaryCIFC}
\end{figure}

In this section, we investigate the dual class of Multi-Primary CIFC referred to as the Multi-Secondary CIFC, as depicted in Fig.~\ref{MultiSecondaryCIFC}. In such a setting, we assume the existence of only one primary receiver but as many as $M$ secondary decoders $Z_k$ with $k \in [1:M]$.  
Similar to the Multi-Primary CIFC, our aim is to characterize the capacity region for several different  interference regimes which will be defined in a somewhat dual manner to the previous section. It is worth mentioning at this point that, due to the \emph{asymmetry} of the roles played by both primary and secondary encoders/receivers, the results of the two sections, though they might exhibit some similarities, originate from fundamentally different challenges to solve as will be later elaborated on. 

\subsection{Capacity region of the Multi-Secondary CIFC in the Very Strong Interference regime}

\begin{assumption}[Very strong interference]
The \emph{very strong interference} regime is defined by: 
\begin{IEEEeqnarray}{rCl}
(\forall k \in [1:M])  \ , \ I(X_2;Z_k | X_1) &\leq& I( X_2; Y | X_1) \ ,  \label{EquStrongIFC}\\
I(X_1 X_2;Y) &\leq& I( X_1X_2;Z_k) \ . 
\end{IEEEeqnarray}
for all joint PDs $P_{X_1X_2}$. 
\end{assumption}

It is worth to mention the above assumption implies that all users $Z_k$ and $Y$ are assumed to experience strong interference, and thus can all of them decode the interference. 

\begin{theorem}[Very strong interference]
The capacity region of the Multi-Secondary CIFC in \emph{very strong interference} is defined by all rate pairs verifying:
\begin{IEEEeqnarray}{rCl} 
R_2 &\leq& \min_{k \in[1:M]} I(X_2; Z_k| X_1) \ ,  \\
R_1 + R_2 &\leq& I(X_1 X_2; Y) \ , 
\end{IEEEeqnarray}
for some joint PD $P_{X_1X_2}$.  
\end{theorem} 
\begin{IEEEproof}
The inner bound follows from superposing the codeword  letting similarly to previously $Q_1 = X_1$, $V_1 = V_2 = V$ and $Q = U = V = X_2$, i.e., allowing all users to decode all codewords.  The proof of converse is standard and follows along similar lines to the proof of Multi-Primary CIFC presented in Appendix~\ref{App-VSI-MultiPrimary}.
 \end{IEEEproof}
 
\subsection{Capacity region of the Multi-Secondary CIFC in the Very Weak Interference Regime}

\begin{assumption}[Very weak interference]
The \emph{very weak interference} regime is defined by 
\begin{IEEEeqnarray}{rCl}
  I(U;Y | X_1) &\leq& \min_{k \in [1:M]}I( U; Z_k| X_1) \ , \label{EquWeakIFC} \\
I(U X_1 ;Y) &\leq& \min_{k \in [1:M]}I( UX_1; Z_k) \ \label{EquVeryWeakIFC} ,
\end{IEEEeqnarray} 
for all $P_{UX_1 X_2}$ verifying $U \mkv (X_1,X_2) \mkv (Y, Z_1 ,\dots, Z_M)$.
\end{assumption}

\begin{theorem}[Very weak interference]
The capacity region of the Multi-Secondary CIFC in \emph{very weak interference} regime is defined by all rate pairs verifying: 
\begin{IEEEeqnarray}{rCl} 
R_1 &\leq&  I(U X_1;Y ) \ ,  \\ 
R_2 &\leq& \min_{k \in[1:M]} I(X_2; Z_k | U X_1) \ , 
\end{IEEEeqnarray}
where $U \mkv (X_1,X_2) \mkv (Y, Z_1 ,\dots, Z_M)$.  
\end{theorem}  
\begin{IEEEproof}
The inner bound follows from letting $Q_1 =(U, X_1)$, $V_1= V_2 = V = X_2$ and using both constraints \eqref{EquWeakIFC} and \eqref{EquVeryWeakIFC}. The converse proof is more challenging, is dual to the proof made for the Multi-Primary CIFC and is presented in Appendix~\ref{App-VWI-MultiSecondary}. 
\end{IEEEproof}

 \subsection{Capacity region of the Multi-Secondary CIFC in the Mixed Very Weak/Strong Interference Regime}

\begin{assumption}[Mixed very weak/strong interference]
Assume the set of secondary users can be partitioned in two subsets: $\mathcal{S}$ corresponding  to the set of users experiencing \emph{very strong interference}, and $\mathcal{W}$ indicating the set of users that experience \emph{very weak interference}. Accordingly, the \emph{mixed very weak/strong interference} regime is defined by the set of inequalities: 
 \begin{IEEEeqnarray}{rCl}
 I(U;Y | X_1) &\leq& \min_{k \in \mathcal{W}}I( U; Z_k| X_1) \ , \label{EquMixedIFC1} \\
I(U X_1 ;Y) &\leq& \min_{k \in \mathcal{W}}I( UX_1; Z_k) \ ,  \label{EquMixedIFC2} \\
(\forall j \in \mathcal{S})  \ , \ I(X_2;Z_k | X_1) &\leq& I( X_2; Y | X_1) \ , \label{EquMixedIFC3} \\
I(X_1 X_2; Y) &\leq& \min_{ k \in \mathcal{S}}I( X_1X_2; Z_k) \  \label{EquMixedIFC4} ,
\end{IEEEeqnarray}
for all joint PDs $P{UX_1X_2}$ such that $U \mkv (X_1,X_2) \mkv (Y, Z_1 ,\dots, Z_M)$. 
\end{assumption}

\begin{theorem}[Mixed very weak/strong interference]
The capacity region of the Multi-Secondary CIFC in \emph{mixed very weak/ strong interference} regime is the set of rate pairs satisfying: 
\begin{IEEEeqnarray}{rCl} 
R_2 &\leq& \min_{k \in \mathcal{S}} I(X_2; Z_k| X_1) \ ,  \\
R_1 + R_2 &\leq& I(X_1 X_2; Y) \ , 
\end{IEEEeqnarray}
for some joint PD $P_{X_1X_2}$.  
\end{theorem}

\begin{IEEEproof}
Converse proof is straightforward. Achievability follows from evaluating the rate region $\mathcal{M}$ with the choice $V_1 = V_2 =\emptyset$, $Q_1 = X_1$, and $Q = U = V = X_2$ which is the optimal scheme for \emph{very strong interference}. 
Then, we note that: 
\begin{IEEEeqnarray}{rCl}
\textrm{Expressions }\ \eqref{EquMixedIFC1} \ \text{ and } \   \eqref{EquMixedIFC3} \ &\Rightarrow& \ \min_{k \in \mathcal{S}}I( X_2; Z_k| X_1)  \leq \min_{k \in \mathcal{W}}I( X_2; Z_k| X_1)  \ , \quad \\
\textrm{Expressions }\ \eqref{EquMixedIFC2}\   \text{ and }  \ \eqref{EquMixedIFC4} \ &\Rightarrow& \ I(X_1 X_2; Y)  \leq  \min_{k \in \mathcal{S} \cup \mathcal{W}}I( X_1X_2; Z_k)  \ , 
\end{IEEEeqnarray}  
which leads to the capacity region formulation.
\end{IEEEproof}

Unlike the Multi-primary CIFC, the inner bound \eqref{CDcoding} is optimal for all discrete memoryless interference regimes for which capacity region was characterized. Yet, the Gaussian counterpart, i.e., the Gaussian Multi-Secondary CIFC, appears to be more challenging, especially in the \emph{weak interference} regime. Hence, it will require the introduction of a more evolved \emph{encoding technique}. 
 
 \subsection{Capacity results for the Gaussian Multi-Secondary CIFC}\label{SectionGaussian}
\begin{figure}[t]
\centering \includegraphics[scale=.6]{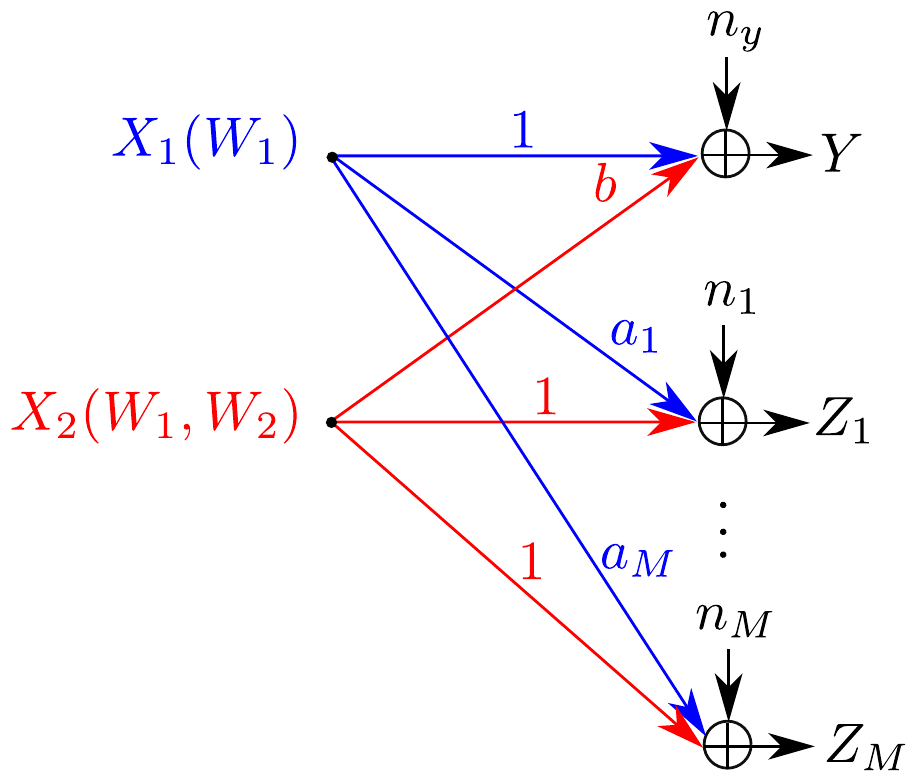}
\caption{The Gaussian Multi-Secondary Cognitive Interference Channel.}
\label{GaussianMultiSecondary}
\end{figure}

Let us consider the following channel model as depicted in Fig. \ref{GaussianMultiSecondary}: 
\begin{IEEEeqnarray}{rCl}
Y &=& X_1 + b X_2 +  N \ , \\ 
Z_1 &=& X_2 + a_1 X_1 + N_1 \ ,\\
Z_2 &=& X_2 + a_2 X_1 + N_2 \ , 
\end{IEEEeqnarray}
where $(N,N_1,N_2)$ are independent and $N,N_1,N_2\sim \mathcal{N}(0,1)$.  
 \subsubsection{The Very Strong Interference regime}
 
 \begin{assumption}[Very strong interference]
 The \emph{very strong interference} regime in the Gaussian setting is defined by the couple of constraints: 
 \begin{IEEEeqnarray}{rCl}
&& |b| > 1 \, , \,\forall \rho \in [-1:1]\  \ , \ \forall k \in [1:2] \ , \nonumber \\
&&  (1-a_k^2)P_1 + (b^2 -1) P_2 + 2 \rho( b - a_k) \sqrt{P_1 P_2} \leq 0  \ . 
 \end{IEEEeqnarray} 
 \end{assumption}
The capacity region in this case is given by the following.
  
 \begin{corollary}[Very strong interference]
 The capacity of the Gaussian Multicast CIFC in the\emph{ very strong interference} regime is given by the set of rate pairs satisfying: 
 \begin{IEEEeqnarray}{rCl}
 R_2 &\leq& \dfrac{1}{2} \log_2 \left( 1 +  \eta P_2\right) \ , \\
 R_1 + R_2 &\leq& \dfrac{1}{2} \log_2 \left( 1 + b^2 P_2 + P_1 + 2 |b|   \sqrt{(1-\eta) P_1 P_2} \right) , 
 \end{IEEEeqnarray}
 for some $\eta \in [0:1]$. 
 \end{corollary}  
\begin{IEEEproof}
 The inner bound follows from noting that, since conditioned on $X_1$ all channels $Z_j$ have equal statistics, the index $j$ does not play a role in the capacity region. The proof follows along similar lines as in \cite{RiniGaussian} since it is equivalent to a non-mutlticast CIFC. 
 \end{IEEEproof}
 \vspace{2mm}
\subsubsection{The Weak Interference regime}
\begin{assumption}[Weak Interference regime]  
The weak interference regime is defined by $| b |  \leq 1 $ which is equivalent to satisfy expression~ \eqref{EquWeakIFC}. 
\end{assumption}
Note here, that we again investigate for the Gaussian case, a \emph{weak interference} regime as opposed to the discrete memoryless case where we can only characterize the capacity region of the \emph{very weak interference} regime. 

Indeed, in the non-multicast setting, $Z_1 = Z_2 = Z$, i.e., $a_1 = a_2 =a $, the optimal strategy is to apply Dirty-Paper Coding by letting the codeword destined to user $Z$, i.e. $V$, precode against interference $(X_u,X_1)$ which is the signal intended to user $Y$. It is only at this condition that we can alleviate the conditions of \emph{very weak interference} to that of  weak interference~\eqref{EquWeakIFC} since the sum-rate decoding constraint is no longer necessary. However, in the multicast setting, by setting $X_2 = X_v + X_u $, we have that: 
\begin{equation}
 Z_k = X_v + \overbrace{X_u}^{\text{Common IFC}} + \overbrace{a_k X_1}^{\text{Private IFC}} + N_k      \ ,
\end{equation}
the transmission to both users $Z_k$ experiences different interference components $X_u + a_1 X_1$ and $X_u + a_2 X_1$, it thus becomes inefficient to apply only a ``common" DPC scheme. 

We remark that a similar challenging setting, though less general,  was investigated in~\cite{Khisti2007} when dealing with simultaneous transmissions to a single-user having uncertainty about two different interferences, e.g. $a_1 S^n$ and $a_2 S^n$, and where tight upper and lower bounds on the capacity region of such a setting are derived, but under the assumption that the sequence $S^n$ is i.i.d and Gaussian distributed. Besides having to deal with similar uncertainty about the interfering sequence, we cannot assume an i.i.d. distribution for the input $X_1$, which renders the analysis of the inner and the outer bounds even more challenging.  

The authors resort here to previous work presented as ``Multiple Description Coding" that was investigated within the framework of the \emph{Compound Broadcast Channel}~\cite{BenammarJournalCompound} but that can be equivalently stated in the Multicast setting as well. In this precoding scheme, we introduce two private descriptions, each decoded by a distinct destination $Z_k$, leading thus to the rate region $ \mathcal{R}$ defined by:    
\begin{IEEEeqnarray}{rCl}\label{MDCodingInnerBound}
R_2 &\leq& \min_{k\in [1:2]} \bigl\{ I( VV_k; Z_k) - I(VV_k;U X_1)  \bigr\} \ , \\ 
2\, R_2 &\leq& \sum_{k\in [1:2]} \bigl[ I( VV_k; Z_j) - I(VV_k;U X_1)\bigl]  - I(V_1; V_2|VUX_1)   \ , \nonumber \\
R_1 &\leq& I(UX_1; Y) \ .    
\end{IEEEeqnarray} 
Our aim is to show that, compared to the inner bound based on ``Common Description" coding:   
\begin{IEEEeqnarray}{rCl}\label{CDCodingInnerBound}
R_2 &\leq& \min_{k \in [1:2]} I( V ; Z_k) - I(V ;U X_1)  \ ,  \\
R_1 &\leq& I(UX_1; Y) \ ,  
\end{IEEEeqnarray}
\emph{multiple description} coding can strictly enhance the rates at users $Z_k$.  To this end, we construct two DPC schemes based on the two different ideas: MD-DPC and CD-DPC, and compare their respective performances. In the sequel, let: \begin{equation}
(X_1 , X_u) \sim \mathcal{N}\left( \mathbf{0} , \Sigma_{X_1X_u} \right)\ \ \textrm{with } \ \  \Sigma_{X_1X_u}=\left(\begin{array}{cc}
P_1 & \rho \sqrt{P_1 P_u} \\ 
\rho \sqrt{P_1 P_u}  & P_u
\end{array} \right)  
\end{equation}
so that $P_u = (1 - \eta) P_2 $ and let $P_v =  \eta P_2 $. Thus,   
\begin{IEEEeqnarray*}{rCl}
R_1  &\leq& I(UX_1;Y) =  \dfrac{1}{2} \log_2 \left( \dfrac{b^2 P_2 + P_1 + 2 b \rho\sqrt{ P_1(1- \eta)P_2}+ 1}{b^2 \eta P_2 + 1} \right)\ . 
\end{IEEEeqnarray*}
\subsubsection{Common Description Dirty-Paper Coding} Consider the CIFC without multicasting, and set the auxiliary random variable: 
\begin{IEEEeqnarray}{rCl}\label{CD-DPC}
V = X_v + \gamma X_u + \alpha X_1 \ . 
\end{IEEEeqnarray}  
The maximal rate $R_2$ achieved by this DPC scheme is then: 
\begin{IEEEeqnarray}{rCl}
 \max_{\gamma, \alpha} \bigl[ I(V;Z) - I(V;X_u X_1) \bigr] &=& I(X_v;Z|X_u X_1) \\
 &=&\dfrac{1}{2} \log_2(P_v+1) \ , 
\end{IEEEeqnarray} 
and is achieved with: $
\gamma_{opt} = \dfrac{P_v}{P_v + 1}  \  \text{and} \   
\alpha_{opt} = a \dfrac{P_v}{P_v + 1} \ . $
However, in presence of a multicast setting, $ \gamma$ cannot be chosen optimal for both channels, which justifies the appellation of common interference. For instance, the private interference cannot be suppressed entirely for both channels at once since it would require that  $\alpha_{opt,1} =  a_1 \dfrac{P_v}{P_v + 1} $ and $\alpha_{opt,2} =  a_2 \dfrac{P_v}{P_v + 1}$.

\begin{corollary}[CD-DPC lower bound]
The maximal rate $R_2$ achieved under the CD-DPC scheme \eqref{CD-DPC} is given by: 
\begin{IEEEeqnarray}{rCl}
\max_{\alpha ,  \gamma} \min_{k\in[1:2]} \bigl[ I(V; Z_k) &-& I(V; UX_1) \bigr] =\frac{1}{2}  \log_2 \left(  P_v+  \right)  \\
  &-&  \frac{1}{2}  \log_2 \left(  \dfrac{P_1 P_v \left(P_v + (1-\rho^2) P_u + 1 \right)(a_1 - a_2)^2}{(P_v + 1)  \left(\sqrt{v_1}+\sqrt{v_2}  \right)^2}  + 1   \right)   \nonumber 
\end{IEEEeqnarray} 
with 
$
 v_k \equiv  \textrm{Var}(Z_k) = P_2 + a_k^2 P_1 + 2 a_k \rho \sqrt{P_1 P_u} + 1 \ .  
$
\end{corollary}
\begin{IEEEproof}
As argued previously, the optimal DPC parameter for the common interference is $\gamma_{opt} =\dfrac{P_v}{(P_v +1)}$ while the optimal $\alpha$ is given by: 
\begin{equation}
 \alpha_{opt,12} = \dfrac{ a_2\sqrt{ v_1} +a_1 \sqrt{v_2}}{\sqrt{v_1}+\sqrt{v_2} } \dfrac{P_v}{P_v + 1}  \ . 
\end{equation} 
The remaining is simple algebra and thus, omitted.
\end{IEEEproof} 

\subsubsection{Multiple Description Dirty-Paper Coding} Using the inner bound $\mathcal{R}$ given in expression~\eqref{MDCodingInnerBound}, we derive a \emph{multiple description} coding based DPC scheme where we let: 
\begin{IEEEeqnarray}{rCl}
V &=& X_v + \gamma  X_u +\alpha X_1 \ , \\
V_k &=& X_p + \gamma_k X_u + \alpha_k X_1 \ , 
\end{IEEEeqnarray} 
in which we split the common description power $P_v$ into $P_v - x$ and $x$ and where: 
$X_v \sim \mathcal{N}(0,P_v- x) \ \bot \  X_p \sim \mathcal{N}(0,x)$. 

We further combine with these DPC codewords a time sharing argument where the two private descriptions are activated in different time slots. This yields a zero correlation cost: 
\begin{equation}
 I(V_1; V_2 | V U X_1 T ) = 0\ . 
\end{equation}  
\begin{corollary}[MD-DPC lower bound]
The optimal private description yields the rate:  
\begin{IEEEeqnarray*}{rCl}
 \max_{\alpha ,  \gamma} \min_{k\in[1:2]} \max_{\alpha_k} \bigl[ I(VV_k; Z_k|T) &-& I(VV_k; UX_1|T) \bigr] =  \frac{1}{2}   \log_2  \left(  P_v+ 1 \right) \\
  &- &  \frac{1}{2}   \log_2 \left( \dfrac{P_1 \left(P_v + (1-\rho^2) P_u + 1 \right)(a_1 - a_2)^2 P(x)}{(P_v + 1) \left(\sqrt{v_1}+\sqrt{v_2}  \right)^2}  + {\sqrt{x+1}}    \right) \nonumber 
\end{IEEEeqnarray*}
where $P(x) \equiv   \dfrac{P_v - x}{\sqrt{x+1}}$. 
\end{corollary}
\begin{IEEEproof} 
The proof is similar to the optimization of the private descriptions in \cite{BenammarJournalCompound} and is thus omitted. 
\end{IEEEproof}

In the sequel, we wish to compare the relative behaviour of both inner bounds through analysing the behaviour of MD-DPC as function of $x$ (the case $x=0$ corresponds to CD-DPC inner bound).  We wish also to evaluate their gap to the best outer bound that we can characterize for such a setting. 

\subsubsection{An outer bound for the weak interference regime} We can write an outer bound for the weak interference regime inspired from the non-multicast CIFC. 
\begin{corollary} [Outer bound for the Gaussian weak interference regime]
The capacity region of the Gaussian Multicast CIFC is included in the set of rate pairs satisfying: 
\begin{IEEEeqnarray}{rCl}
 R_1 &\leq& \dfrac{1}{2} \log_2 \left( \dfrac{b^2 P_2 + P_1 +  2  | b | \sqrt{(1-\eta)P_1 P_2}+ 1}{b^2 \eta P_2 + 1} \right)  \ , \quad \\ 
 R_2 &\leq&  \dfrac{1}{2} \log_2 \left(  \eta P_2 + 1  \right) \ , 
 \end{IEEEeqnarray}
 for some arbitrary $\eta \in [0:1]$.   
\end{corollary}
 \begin{IEEEproof}
  The proof follows along the same lines as in \cite{Wu2007} and results from Fano's inequality that implies: 
 \begin{IEEEeqnarray}{rCl}
 nR_2 &\leq& I(X_2^n; Z_k^n | W_1 X_1^n) + n \epsilon_n  \ , \\
    nR_1 &\leq& I(W_1 X_1^n; Y^n) + n \epsilon_n \ ,
 \end{IEEEeqnarray} 
 which turns out to be a non-multicast setup since conditioning on $X_1$, all channels $Z_k$ share the same statistics. 
 \end{IEEEproof} 
 
\subsubsection{Comparison of the inner and outer bounds} 
\begin{figure}[t]  
 \centering\includegraphics[scale=0.7]{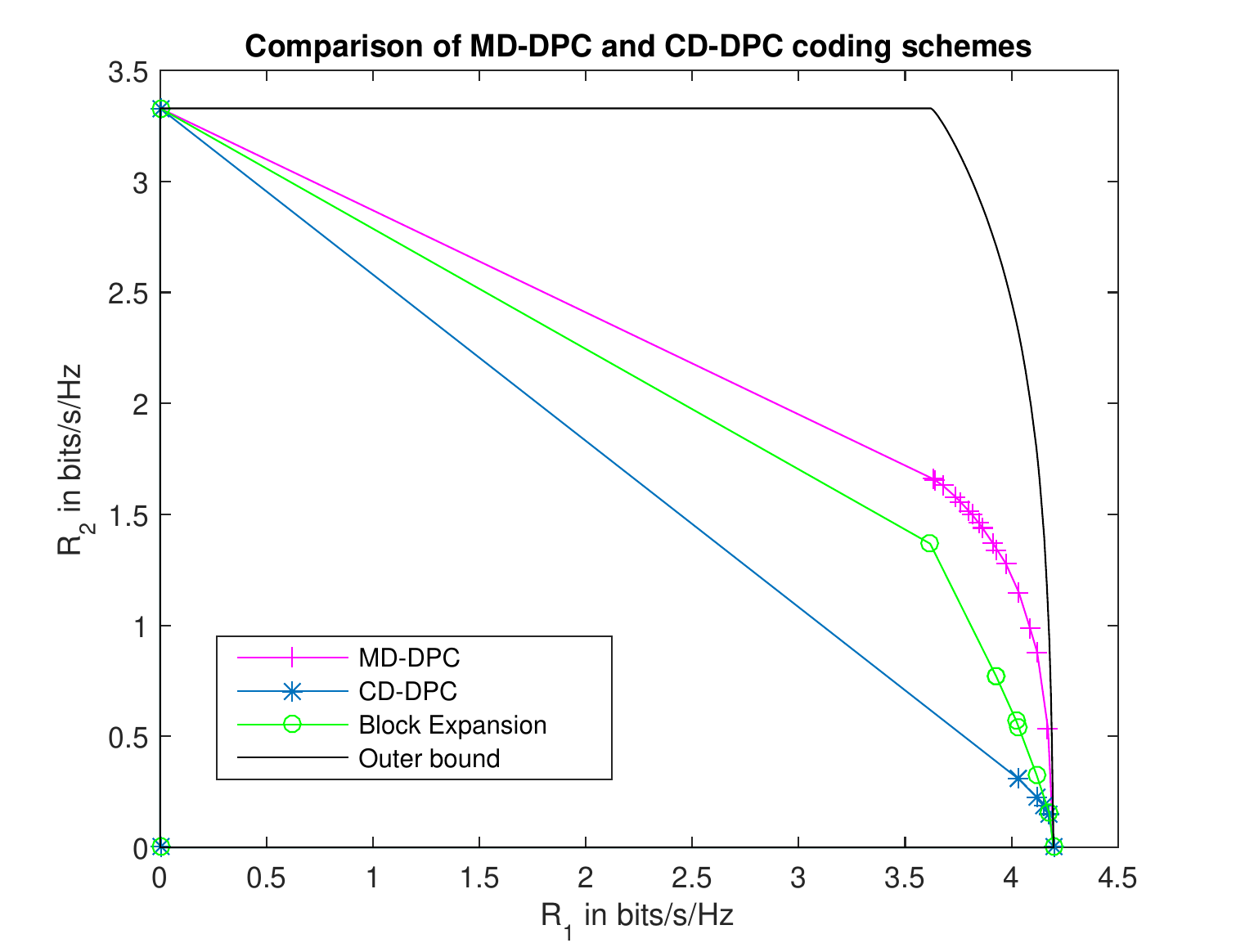} 
 \caption{Comparison of the bounds with: $P_1 = 3 P_2$, $a_1 = .75$, $a_2 = - .5$, $b= .1$.}
 \label{ComparisonBounds} 
 \end{figure}

In the light of the results in~\cite{BenammarJournalCompound}, we can show by a simple function study that, the rate achieved by \emph{multiple description} coding can be strictly increasing in $x$ and thus $x=0$ (CD-DPC) yields its minimum value. 
In Fig. \ref{ComparisonBounds}, we plot the different inner bounds where we further combine the CD-DPC to a time-sharing argument, referred to as \emph{block expansion} in~\cite{WeingartenDoF}. It is clear then that the improved endowed by \emph{multiple description} coding is strict and that it is becomes more relevant when the mismatch between the multicast channels $|a_1 - a_2|$ and the primary user power $P_1$ increase. 
  
We investigated the Multi-Secondary CIFC for which we characterize the capacity region in some interference regimes and show that, for the Gaussian \emph{weak interference} case, a more evolved encoding scheme than the one allowed by the inner bound~\eqref{CDcoding}, is required to enhance the transmission rates.
   
 %
 %
   \section{Comments and Discussion } \label{Sec-Vaezi}
  
  \subsection*{Anecdotic result}
In a recent paper by Vaezi~\cite{VaeziWISI}, it is shown that the \emph{better cognitive decoding} (BCD) introduced in \cite{Rini2011} is nothing but the mere \emph{ very weak interference} (VWI) regime and that the new capacity result in this regime is an equivalent formulation of the VWI capacity region. However, the astonishing result in the work of Vaezi lies in proving that the \emph{very strong interference} (VSI) regime is contained in the VWI regime for finite alphabets suggesting that apparently  contradictory regimes from a conceptual point of view, are in fact equivalent. Later on, we clarify how the present work does not fall in such a \emph{triviality} be it even for finite alphabet settings.  

A crucial remark here is that the claim of Vaezi that the class of probability distribution that verify (VSI) falls into the class of (VWI) or equivalently, the (BCD), does not hold in the multicast setting. Let us first review this claim of \emph{triviality} by  Vaezi. The three regimes are given by the following conditions: 
  \begin{IEEEeqnarray}{rCl} 
    \text{(VWI)}  \ ,  \ \forall P_{UX_1X_2}  &:& 
    \left\{\begin{array}{rcl}
      I(U ;Y|X_1) &\leq& I(U ;Z|X_1)\ , \\
      I( X_1;Y) &\leq& I( X_1;Z)\ ,
    \end{array}\right. \\
    \text{(BCD)} \ , \ \forall P_{UX_1X_2} &:&  \ I(UX_1;Y) \leq I(UX_1;Z) \ ,\\
    \text{(VSI)} \ , \ \forall P_{X_1X_2}  &:& 
    \left\{\begin{array}{rcl }
     I(X_1X_2;Y) &\leq& I(X_1X_2;Z)\ , \\
      I(X_2;Z | X_1) &\leq& I(X_2;Y | X_1) \ . 
\end{array}\right.    
  \end{IEEEeqnarray}  
  The key tool to show the equivalence between VWI and BCD is to notice that: 
  \begin{equation}
    \forall P_{UX_1X_2} \ , \ I(U X_1 ;Y) \leq I(UX_1 ;Z ) \Rightarrow  \forall P_{UX_1X_2} \ , \ I(U ;Y|X_1) \leq I(U ;Z|X_1) \ .  
   \end{equation}  
   Thus, since: 
   \begin{eqnarray}
    I(X_1X_2;Y) \leq I(X_1X_2;Z) \ \text{and} \ I(X_2;Z | X_1) \leq I(X_2;Y | X_1)  \\ 
      \Rightarrow \forall P_{U} \ , \ I(UX_1;Y) \leq I(UX_1;Z)  \ ,  
   \end{eqnarray} 
   which means that VSI implies BCD which in turn is equivalent to VWI. 
   
   However, e.g. in the multi-primary setting, the distinct regimes are defined by: 
     \begin{IEEEeqnarray}{rCl} 
    \text{(VWI)}  \ ,  \ \forall P_{UX_1X_2}  &:& 
    \left\{\begin{array}{rcl}
     \forall j \in[1:N] \ ,\quad  I(U ;Y_j|X_1) &\leq& I(U ;Z|X_1) \ ,\\
      \displaystyle\min_{j \in [1:N]} I(UX_1;Y_j) &\leq& I(UX_1;Z)\ ,
    \end{array}\right.  \\ 
    \text{(VSI)} \ , \ \forall P_{X_1X_2}  &:& 
    \left\{\begin{array}{rcl }
      \forall j\in[1:N] \ ,  \quad   I(X_2;Z | X_1) &\leq& I(X_2;Y_j | X_1)\ , \\
      \displaystyle\min_{j\in[1:N]} I(X_1X_2;Y_j) &\leq& I(X_1X_2;Z) \ . 
\end{array}\right.    
  \end{IEEEeqnarray}  
  It is clear when there is only one primary channel output, i.e., $N=1$, that what we recover is the BCD and the VSI regime. However, when $N> 1$, there is no evidence why the VSI should be included in the VWI since: $VSI$ implies $\min_{j\in[1:N]} I(UX_1;Y_j) \leq  I(UX_1 ;Z)$,  but the other constraint of VWI can not be implied since it is too strict:
   \begin{equation}
    \text{VSI}  \ \nRightarrow\ , \ \forall j \in[1:N] \quad  I(U ;Y_j|X_1)  \leq  I(U ;Z|X_1)  \ . 
   \end{equation}
   Thus, VSI cannot imply VWI for all classes of multicast CIFC. 
   
   \subsection{Future line of work}
The analysis we made in this paper, for the Multi-Primary CIFC and Multi-Secondary CIFC, would serve as a baseline to further study  the Multicast CIFC with arbitrary numbers of primary and secondary users. We emphasize here, however, that this constitutes a challenging task, for two main reasons: one that arises from the fact that the Multi-cast CIFC is even more general than the Compound Broadcast channel, for which only Degrees of Freedom are well understood to date, and the other one that stems from the fact that defining more general interference regimes becomes quickly intractable due to the multiple combinations of users. 
   
Yet, introducing the ideas of Interference Decoding and Multiple Description coding will clearly be of crucial importance when dealing with more general Multicast CIFC settings and thus, a thoughtful combination of the two schemes seems to be needed 
 for such settings.

 %
 %
 \appendices 
 
 \section{Basic properties and results}\label{sec:typical}
This appendix provides basic notions on some concepts used through this paper. Following definition in~\cite{csiszar1982information}, we use in this paper \emph{strongly typical sets} and the so-called \emph{delta-convention}. 
Some useful facts are recalled here.

\begin{definition}
For any sequence $x^n\in\cX^n$ and any symbol $a\in\cX$, notation $N(a|x^n)$ stands for the number of occurrences of $a$ in $x^n$.
\end{definition}

\begin{definition}
A sequence $x^n\in\cX^n$ is called \emph{(strongly) $\delta$-typical} w.r.t. $X$ (or simply \emph{typical} if the context is clear) if
\[
\left| \frac1n N(a|x^n) - P_X(a) \right| \leq \delta\ , \ \text{ for each } a\in\cX \ ,
\]
and $N(a|x^n)=0$ for each $a\in\cX$ such that $P_X(a)=0$.
The set of all such sequences is denoted by $\typ{X}$.
\end{definition}

\begin{definition}
Let $x^n\in\cX^n$.
A sequence $y^n\in\cY^n$ is called \emph{(strongly) $\delta$-typical (w.r.t. $Y$) given $x^n$}  if
\[
\left| \frac1n N(a,b|x^n,y^n) - \frac1n N(a|x^n)P_{Y|X}(b|a)\right|  \leq \delta \ \text{ for each } a\in\cX, b\in\cY \ ,
\]
and, $N(a,b|x^n,y^n)=0$ for each $a\in\cX$, $b\in\cY$ such that $P_{Y|X}(b|a)=0$.
The set of all such sequences is denoted by $\typ{Y|x^n}$.
\end{definition}

\emph{Delta-convention~\cite{csiszar1982information}:\ }
For any sets $\cX$, $\cY$, there exists a sequence $\{\delta_n\}_{n\in\bN^*}$ such that lemmas below hold.\footnote{As a matter of fact, $\delta_n\to0$ and $\sqrt{n}\,\delta_n\to\infty$ as $n\to\infty$.}
From now on, typical sequences are understood with $\delta=\delta_n$. 
Typical sets are still denoted by $\typ{\cdot}$.

\begin{lemma}[{\cite[Lemma~1.2.12]{csiszar1982information}}]\label{lem-concentration}
There exists a sequence $\eta_n\toas{n\to\infty}0$ such that
\[
P_X^n(\typ{X}) \geq 1 - \eta_n \ .
\]
\end{lemma}

\begin{lemma}[Joint typicality lemma~\cite{cover2006elements}]
\label{lem:jointTypicality}
There exists a sequence $\eta_n\toas{n\to\infty}0$ such that
\[
\left| -\frac1n \log P^n_Y(\typ{Y|x^n}) - I(X;Y)\right|  \leq \eta_n \ ,\
\textrm{ for each }x^n\in\typ{X}\ .
\]
\end{lemma}

\begin{lemma}[{Csisz\'ar \& K\"orner's sum-identity~\cite[Lemma~7]{CsiszarConfidential}}]
Consider two random sequences $X^n$ and $Y^n$ and a random variable $C$.
The following identity holds:
\begin{equation}
\sum_{i=1}^n I(Y_{i+1}^n ; X_i | C X^{i-1})
	=	\sum_{j=1}^n I(X^{j-1} ; Y_j | C Y_{j+1}^n)	\ .\label{sec:csiszarkorner}
\end{equation}
\end{lemma}
 
  \section{Proof of achievability for the Multicast CIFC}\label{ProofOfAchievabilityMulticast}
All the channel outputs in the Multicast setting are to be treated in a similar manner, thus, we show the achievability only with one primary channel output $Y$.  
 
 \underline{\textit{Rate splitting:}}
 
 The message $W_1$ is split in two parts $W_{01}$ of rate $R_{01}$ that is intended to be decoded by all users and a private part $W_{11}$ that is intended  to be decoded only by user $Y$. 

  \underline{\textit{Codebook generation:}} At source $1$: First generate $2^{n R_{01}}$ sequences $q_1^n(w_{01})$ following \\ $\prod_{i=1} ^n P_{Q_1} ( q_{1,i} (w_{01}))$ . For each sequence $q_1^n(w_{01})$, generate $2^{n R_{11}}$ sequences $x_1^n(w_{11})$ following $\prod_{i=1} ^n P_{X_1|Q_1} ( x_{1,i} (w_{11},w_{01}))$. 
  
  At Source $2$: For each sequence $q_1^n(w_{01})$, generate $ 2^{n( T_{02} )}$ sequences $q^n( w_{01},s_{02})$ following $ \prod_{i=1} ^n P_{Q|Q_1} ( q_{i} (w_{01},s_{02}))$ and throw them in $ 2^{n  R_{02} }$ bins $B_0^n( w_{01},w_{02})$. For each sequence $q^n( w_{01},s_{02})$, generate $2^{n T_{11}}$ sequences $u^n( s_{11},w_{01},s_{02})$ and $2^{n T_{22}}$ sequences $v^n( s_{22},w_{01},s_{02})$ following $\prod_{i=1} ^n P_{U|QQ_1} ( u_{i} (s_{11}, w_{01},s_{02}))$  and  $\prod_{i=1} ^n P_{V|QQ_1} ( v_{i} (s_{22}, w_{01},s_{02}))$ and throw them respectively in $2^{n R_{11} }$ bins $B_1^n( w_{11},w_{01},s_{02})$ and  $2^{n R_{22} }$ bins $B_2^n( w_{22},w_{01},s_{02})$. 

\underline{\textit{Encoding:}}  The encoder $1$ sends $x^n_{1} (w_{11},w_{01})$. The encoder $2$ finds in bin $B_0^n( w_{01},w_{02})$ a sequence indexed by $ s_{02} $ such that: 
\begin{equation}
 \left( x_1^n(w_{11},w_{01}), q_1^n(w_{01} ) , q^n( w_{01}, s_{02}) \right) \in \mathcal{T}^{(n)}_{[QQ_1X_1]} \ . 
\end{equation}
 Then, it looks in the product bin $ B_1^n( w_{22},w_{01},s_{02})\times B_2^n( w_{22},w_{01},s_{02}) $ for a couple of sequences such that: 
\begin{equation}
 \left( x_1^n(w_{11},w_{01}), q_1^n(w_{01} ) , q^n( w_{01}, s_{02}) , u^n( s_{11},w_{01},s_{02}),  v^n( s_{22},w_{01},s_{02}) \right) \in \mathcal{T}^{(n)}_{[QQ_1UVX_1]} \ .  
\end{equation}
 It then sends a codeword $x^n_2 ( s_{11},s_{22} ,w_{01},s_{02})$. 
 
 The encoding will be flawless if the following inequalities hold: 
  \begin{IEEEeqnarray}{rCl} 
 T_{02} - R_{02}  &\geq& I(X_1;Q |Q_1)  \ , \\ 
 T_{11} - R_{11} &\geq& I(U; X_1 |Q_1Q ) \ ,  \\ 
 T_{22} - R_{22} &\geq& I(V; X_1 |Q_1Q )  \ ,  \\ 
 T_{11} - R_{11} + T_{22} - R_{22} &\geq& I(U; V |Q_1Q )  + I(UV; X_1 |Q_1Q )   \ . 
 \end{IEEEeqnarray}

\underline{ \textit{Decoding:}}  Receiver $2$ decodes simultaneously the indices $(w_{01}, s_{02} , s_{22})$ while decoder $1$ decodes simultaneously $ (w_{01},s_{02},w_{11},s_{11})$. The probability of error can be made arbitrarily small if the following inequalities hold: 
  \begin{IEEEeqnarray}{rCl} 
 T_{22} &\leq& I(V; Z|QQ_1)  \ , \\ 
 T_{02} + T_{22} &\leq& I( Q V; Z|Q_1)  \ , \\ 
 R_{01} + T_{02} + T_{22} &\leq& I( Q_1 Q V; Z )  \ , \\ 
 T_{11} &\leq& I(X_1 U; Y |Q)  +  I( QU;X_1|Q_1)  \ ,  \\
 T_{02} + T_{11} &\leq& I(X_1 Q U; Y|Q_1  ) +  I( QU;X_1|Q_1)  \ ,  \\
 R_{01}+ T_{02} + T_{11} &\leq& I(Q_1 X_1 Q U; Y ) +  I( QU;X_1|Q_1)  \ . 
 \end{IEEEeqnarray}
 
\underline{\textit{Fourrier Motzkin Elimination:}} After running FME on binning rates $ T_{02}, T_{11}, T_{22}$  and on rate splitting parameters $R_{01}$ and $ R_{02}$, we end up with the rate region in Theorem 1.

\underline{\textit{Multicast setting:}} All channel output $Y_j \ , \ j \in [1:N]$ perform all the same decoding strategy, thus, the rate region can be written similarly obtained replacing $Y$ with the minimum over all channel outputs $Y_j$, where $ j \in [1:N]$. The same applied for all users $Z_k$, $k \in[1:M]$. 

\section{Multi-Primary Very Strong Interference capacity region}\label{App-VSI-MultiPrimary}
  The proof of converse follows from that the strong interference condition \eqref{StrongIFcondition} implies that for all $P_U$ s.t $U\mkv (X_1,X_2) \mkv (Z, Y_1,\cdots, Y_N)$: 
  \begin{equation} \label{IneqStrongIC}
    \forall j\in [1:N]\ , \qquad I(X_2^n ; Z^n | X_1^n U) \leq  I(X_2^n ;  Y_j^n| X_1^n U)  \ .
   \end{equation} 
   To see this, note the following: 
   \begin{IEEEeqnarray}{rCl}
 I(X_2^n ; Z^n | X_1^n U) &-&  I(X_2^n ;  Y_j^n| X_1^n U) \nonumber \\
   &=& \sum_{i=1}^n \bigl[  I(X_2^n ; Z_i |Z^{i-1}  X_1^n U)  -  I(X_2^n ;  Y_{j,i}|  Y_{j,i+1}^n X_1^n U) \bigr] \\
   &\overset{(a)}{=}& \sum_{i=1}^n \bigl[  I(X_2^n ; Z_i |Z^{i-1} Y_{j,i+1}^n X_1^n U)  -  I(X_2^n ;  Y_{j,i}| Z^{i-1} Y_{j,i+1}^n X_1^n U) \bigr] \\
   &\overset{(b)}{=}& \sum_{i=1}^n \bigl[  I(X_{2,i} ; Z_i |Z^{i-1} Y_{j,i+1}^n X_1^n U)  -  I(X_{2,i} ;  Y_{j,i}| Z^{i-1} Y_{j,i+1}^n X_1^n U) \bigr] \\
   &\overset{(c)}{\geq}& 0  \ , 
   \end{IEEEeqnarray}
   where $(a)$ is a consequence of Csisz\'ar \& K\"orner sum-identity applied twice as follows: 
   \begin{IEEEeqnarray}{rCl} 
   && \sum_{i=1}^n \bigl[  I(X_2^n ; Z_i |Z^{i-1}  X_1^n U)  -  I(X_2^n ;  Y_{j,i}|  Y_{j,i+1}^n X_1^n U) \bigr] \nonumber \\
   &=& \sum_{i=1}^n \bigl[  I(Y_{j,i+1}^n X_2^n ; Z_i |Z^{i-1}  X_1^n U)  -  I(Z^{i-1} X_2^n ;  Y_{j,i}|  Y_{j,i+1}^n X_1^n U) \bigr] \\
   &=& \sum_{i=1}^n \bigl[  I(X_2^n ; Z_i |Z^{i-1} Y_{j,i+1}^n  X_1^n U)  -  I(X_2^n ;  Y_{j,i}| Z^{i-1}  Y_{j,i+1}^n X_1^n U) \bigr]
   \end{IEEEeqnarray}
   and $(b)$ is a consequence of the following Markov Chain: 
   \begin{equation}
    (X_2^{i-1}, X_{2,i+1}^n) \mkv (X_{1,i}, X_{2,i}, X_1^{i-1},X_{1,i+1}^n, Z^{i-1}, Y_{j,i+1}^n) \mkv (Y_{j,i},  Z_{j,i} ) 
   \end{equation}
   for all $i=[1:n]$, and finally $(c)$ is a consequence of the strong interference condition  \eqref{StrongIFcondition}.
   
  Thus, one can write that: 
  \begin{IEEEeqnarray}{rCl} 
  n  ( R_2 - \epsilon_n ) &\leq& I(W_2; Z^n | W_1)    \\
  &\leq& I(X_2^n W_2; Z^n | X_1^n W_1)  \\
  &\leq& \sum_{i=1}^n I(X_2^n W_2; Z_i | Z^{i-1} X_1^n W_1) \\
  &=& \sum_{i=1}^n \Bigl[  H( Z_i | Z^{i-1} X_1^n W_1)- H(  Z_i | Z^{i-1} X_2^nX_1^n W_2 W_1) \Bigr]  \\
  &\leq& \sum_{i=1}^n \Bigl[ H( Z_i | X_{1,i} )- H( Z_i | X_{2,i}X_{1,i} ) \Bigr]  \\
  &=& \sum_{i=1}^n I(  X_{2,i} ; Z_i | X_{1,i} )    \ . 
  \end{IEEEeqnarray}
  Moreover, we can also write a sum-rate: 
   \begin{IEEEeqnarray}{rCl} 
  n (R_1+ R_2 - \epsilon_n ) &\leq& I(W_1; Y^n_j) + I(W_2; Z^n | W_1)    \\
  &\leq& I(W_1  X_1^n; Y^n_j) + I(X_2^n ; Z^n | X_1^n W_1)  \\
  &\overset{(a)}{\leq}&   I(W_1  X_1^n; Y^n_j) + I(X_2^n ;  Y^n_j | X_1^n W_1)   \\ 
  &=& \sum_{i=1}^n I( X_{1,i} X_{2,i} ; Y_{j,i}  )   \ , 
  \end{IEEEeqnarray}
  where $(a)$ is a consequence of inequality~\eqref{IneqStrongIC}. The proof of the outer bound follows as usually by defining a time-sharing random variable $Q$ taking values on $[1:n]$ with uniform probability.

  \section{Multi-Primary Very Weak Interference capacity region}\label{App-VWI-MultiPrimary}
  Consider the N-multicast CIFC under the weak interference condition given by~\eqref{WeakInterferenceCondition}. In order to show the converse to the region given by
   \begin{equation}
   \left\{ \begin{array}{rcl}
     R_1 &\leq& \displaystyle\min_{j\in [1:N]}  I(UX_1;Y_j)    \ , \\
   R_2 &\leq& I(X_2;Z|X_1U)  \ , 
   \end{array}\right.
   \end{equation} 
  for some arbitrarily dependent variables $(U,X_1,X_2)$ s.t $U\mkv (X_1,X_2) \mkv (Z, Y_1,\cdots, Y_N)$,  let $j\in [1:N]$. We have that from Fano's inequality:
    \begin{IEEEeqnarray}{rCl} 
   n ( R_2  -  \epsilon_n ) &\leq& I(W_2; Z^n | W_1 X^n_1) \\
   &\leq& I(X_2^n; Z^n | W_1 X^n_1) \\
   &=& \sum_{i=1}^n I(X_2^n; Z_i |Z^{i-1} W_1 X^n_1) \\
   &\leq&  \sum_{i=1}^n I(X_{2,i}; Z_i |U_i X_{1,i}) \ , 
   \end{IEEEeqnarray}
   where we defined $U_i \equiv( W_1, X_1^{i-1}, X_{1,i+1}^n ,Z^{i-1})$. Now on the other side, 
   \begin{IEEEeqnarray}{rCl}
   n ( R_1 -   \epsilon_n )  &\leq& I(W_1 X^n_1 ; Y_j^n) \\
   &=& \sum_{i=1}^n I( W_1 X^n_1 Y_j^{i-1};  Y_{j,i} ) \ .  
   \end{IEEEeqnarray}
   The main idea here is that, we can upper bound this expression letting $ Z^{i-1} $ replace  $Y_j^{i-1}$, i.e., 
   \begin{equation}
 \forall i \in [1:n] \quad , \quad I( W_1 X^n_1 Y_j^{i-1};  Y_{j,i} ) \leq  I( W_1 X^n_1 Z^{i-1};  Y_{j,i} )  \ .
\end{equation}
    This is due to the conditional less-noisiness of $Y_j$ compared to $Z$ given $X_1$ in~\eqref{WeakInterferenceCondition} and it is easy to check that by following similar lines as in~\cite{NairLessNoisy}: 
      \begin{IEEEeqnarray}{rCl} 
I( W_1 X^n_1 Z^{i-1};  Y_{j,i} ) &-&  I( W_1 X^n_1 Y_j^{i-1};  Y_{j,i} ) \nonumber \\
    &=& I(Z^{i-1};  Y_{j,i} | W_1 X^n_1  )  -  I(Y_j^{i-1} ;  Y_{j,i} |  W_1 X^n_1)   \\
    &=& \sum_{r= 1}^{i-1}  \Bigl[ I( Y_j^{r-1} Z_{r}^{i-1} ; Y_{j,i} |   W_1 X^n_1 ) - I( Y_j^{r} Z_{r+1}^{i-1} ; Y_{j,i} |   W_1 X^n_1 ) \Bigr] \\
    &=& \sum_{r=1 }^{i-1} \Bigl[I( Z_{r}  ; Y_{j,i} |   Y_j^{r-1} Z_{r+1}^{i-1}  W_1 X^n_1 ) - I( Y_{j,r}  ; Y_{j,i} |   Y_j^{r-1} Z_{r+1}^{i-1}  W_1 X^n_1 )\Bigr]\  \\
    &=& \sum_{r=1 }^{i-1} \Bigl[ I( Z_{r}  ; Y_{j,i} | X_{1,r}  Y_j^{r-1} Z_{r+1}^{i-1}  W_1  X^n_{1,r+1} X_1^{r-1} ) \nonumber \\
    &-& I( Y_{j,r}  ; Y_{j,i} |   X_{1,r} Y_j^{r-1} Z_{r+1}^{i-1}  W_1  X^n_{1,r+1} X_1^{r-1}) \nonumber  \Bigr] \\
    &\overset{(a)}{\geq}& 0   \ . 
   \end{IEEEeqnarray}
   The condition in \eqref{WeakInterferenceCondition} implies that for all $r\in [1:i-1]$ and all $V $:
    \begin{equation}
 I(U ; Y_r | V X_{1,r})  \leq I(U ; Z_r | V  X_{1,r}) \ . 
\end{equation}
    Letting $U  \equiv Y_{j,i}$  and $V  \equiv (Y_j^{r-1},Z_{r+1}^{i-1} , W_1 ,X^n_{1,r+1} ,X_1^{r-1})$, the claim in $(a)$ is proved. Thus, 
    \begin{IEEEeqnarray}{rCl} 
   n R_1 -  n \, \epsilon_n  &\leq&  \sum_{i=1}^n I( W_1 X^n_1 Y_j^{i-1};  Y_{j,i} ) \\
   &\leq&  \sum_{i=1}^n I( W_1 X^n_1 Z^{i-1};  Y_{j,i} )  \ , 
   \end{IEEEeqnarray}
   which completes the proof. 
  \section{Multi-Primary Mixed weak/strong interference capacity region}\label{App-Mixed-MultiPrimary}
 
For the converse part, we can write that: 
      \begin{IEEEeqnarray}{rCl}
   n R_1 &\leq& \displaystyle \min_{j \in \mathcal{W}} \sum_{i=1}^n I(U_i X_{1,i} ;Y_{j,i}) \ ,  \\
    n R_2 &\leq& \sum_{i=1}^n  I(X_{2,i}; Z_i | U_i X_{1,i}) \ ,  \\
   n (R_1 + R_2 ) & \leq &  \displaystyle\min_{j \in \mathcal{S}} \sum_{i=1}^n  I(X_{1,i}X_{2,i};Y_{j,i}) \ ,
   \end{IEEEeqnarray}
   where $U_i =( W_1, X_{1,i+1}^n, X_1^{i-1}, Z^{i-1})$ for all $i=[1:n]$. 
   
   As for the achievability part, we consider the codebook construction in Appendix~\ref{ProofOfAchievabilityMulticast}, and  let $   Q_1 = \emptyset$ and $Q = U$. We can summarize the encoding constraints as follows: 
     \begin{IEEEeqnarray}{rCl}  
 T_{1} - R_{1} &\geq& I(U; X_1 ) \ ,  \\ 
 T_{2} - R_{2} &\geq& I(V; X_1| U)  \ .    
 \end{IEEEeqnarray}
 
 As for the decoding constraints, user $Z$ decodes the signal $U$ and $X_1$ non-uniquely, finding the unique $s_2$ such that for some $w_1$ and $s_1$: 
   \begin{equation}
    \bigl( u^n(s_1) , x^n_1(w_1), v^n(s_1,s_2), y_j^n \bigr) \in \typ{UX_1VZ}\ ,
   \end{equation}
   where $ u^n(s_1)$ is in the bin defined by $s_1$.
 Thus, we end up with the constraints: 
  \begin{IEEEeqnarray}{rCl}   
 T_{2} &\leq& I(V; Z | X_1 U ) + I(V; X_1| U)  \ ,  \\ 
 T_{1} + T_{2} &\leq& I( X_1 U V ;Z) + I(U V; X_1)   \ . 
 \end{IEEEeqnarray}
 
 On the other side, the users $Y_j$ can choose between two decoding strategies: 
  \begin{itemize}
  \item Not decoding interference, i.e., finding the unique $w_1$ for which: 
  \begin{equation}
 \bigl( u^n(s_1) , x^n_1(w_1), y_j^n \bigr) \in \typ{UX_1Y_j}  \ ,   
  \end{equation}
  where $u^n(s_1) \in B_1^n(w_1)$ is in the bin defined by $w_1$. This yields the following constraint: 
   \begin{equation}
    T_{1} \leq I(U X_1; Y_j) + I(U;X_1)  \ .
\end{equation}     
  \item Decoding interference non-uniquely, finding the unique $w_1$ such that for some $s_2$,
\begin{equation}
   \bigl( u^n(s_1) , x^n_1(w_1), v^n(s_1,s_2), y_j^n \bigr) \in \typ{UX_1VY_j}  \ .  
  \end{equation}  
  This results in the constraint: 
  \begin{equation}
    T_{1} + T_{2} \leq I(U X_1 V; Y_j) + I(VU;X_1) \ .  
\end{equation}   
  \end{itemize}
  One then can write an achievable inner bound with all possible  combinations of decoding choices of each of the users $Y_j$. 
  
Using this idea, we let the group of users in strong interference decode interference as well, and we let the users in weak interference decode only their intended signals $U$ and $X_1$. The resulting set of constraints is given by: 
   \begin{equation}
   \left\{ \begin{array}{rcl}
   T_{1} &\leq& \displaystyle \min_{j \in \mathcal{W}} I(U X_1; Y_j) + I(U;X_1) \ , \\  
   T_{2} &\leq& I(V; Z | X_1 U ) + I(V; X_1| U)  \ ,  \\ 
 T_{1} + T_{2} &\leq& I( X_1 U V ;Z) + I(U V;X_1)   \ , \\
    T_{1} + T_{2} &\leq& \displaystyle\min_{j \in \mathcal{S}}  I(U X_1 V; Y_j) + I(UV;X_1) \ .
   \end{array} \right.
   \end{equation}
 Running FME on the resulting rate region yields: 
    \begin{equation}
   \left\{ \begin{array}{rcl}
   R_1  &\leq& \displaystyle \min_{j \in \mathcal{W}} I(U X_1; Y_j)   \ , \\  
   R_2 &\leq& I(V; Z | X_1 U ) \ ,  \\ 
   R_1 + R_2  &\leq& I( X_1 U V ;Z)    \ , \\
   R_1 + R_2 &\leq& \displaystyle\min_{j \in \mathcal{S}}  I(U X_1 V; Y_j)   \ .
   \end{array}\right.
   \end{equation}
   Letting $V = X_2$, we end up with the following achievable region: 
       \begin{equation}
   \left\{ \begin{array}{rcl}
   R_1  &\leq& \displaystyle \min_{j \in \mathcal{W}}  I(U X_1; Y_j)   \ , \\  
   R_2 &\leq& I(X_2; Z | X_1 U ) \ ,  \\ 
   R_1 + R_2 &\leq& I( X_1 X_2 ;Z)    \ , \\
   R_1 + R_2 &\leq& \displaystyle\min_{j \in \mathcal{S}}  I(X_1 X_2; Y_j)   
   \end{array}\right.
   \end{equation}
   with the strong interference or the weak interference condition, we can show that the sum-rate $R_1 + R_2 \leq I( X_1 X_2 ;Z)$ is redundant. This completes the proof of achievability.

 \section{Proof of Theorem \ref{TheoremGaussianMultiPrimaryCIFC}: Gaussian Multi-Primary CIFC}\label{App-Gaussian-MultiPrimary}
 
 \subsection{Very Strong Interference}
 We start with the achievability part. Consider the following coding scheme: 
\begin{equation}
(X_1,X_2) \sim \mathcal{N} \left( \mathbf{0} ,  \Sigma_{X_1X_2} \right)\ , \ \Sigma_{X_1X_2}=\left[ \begin{array}{cc}
P_1 &  \rho \sqrt{ P_1  P_2} \\ 
 \rho \sqrt{ P_1  P_2} &   P_2 
\end{array} \right] \ . 
\end{equation}
Then, letting $j\in [1:N]$: 
\begin{eqnarray}
I(X_1 X_2; Y_j)& =& \dfrac{1}{2} \log_2\left( 1 + b_j^2P_2 + P_1 + 2 b_j \rho \sqrt{P_1P_2} \right) \ ,  \\
I(X_2;Z |X_1)&=& \dfrac{1}{2} \log_2\left( 1 + (1 - \rho^2)P_2 \right) \ ,  
\end{eqnarray}
 which completes the proof of achievability. 
 
 As for the converse, following the same lines as the proof of the outer bound  of Rini \textit{et al.} \cite{RiniGaussian}, we can write the following outer bound as: 
 \begin{equation}
 \left\{\begin{array}{rcl}
 R_1 &\leq& \displaystyle\min_{j \in [1:N]} I(X_1X_2; Y_j)   \ ,  \\ 
R_2 &\leq& I(X_2; Z|X_1) \ , \\
R_1 + R_2 &\leq&  \displaystyle\min_{j \in [1:N]} I(X_1X_2; Y_j) + I(X_2; Z| Y^\prime_j X_1)  \ . 
 \end{array}\right.
 \end{equation}
  Similarly to the result of \cite{RiniGaussian}, we compute the optimal correlation coefficient between $Z$ and $Y^\prime_j$ conditioned on $X_1$. We obtain then, the following outer bound: 
\begin{IEEEeqnarray}{rCl}
R_2 &\leq&  \dfrac{1}{2} \log_2 \left(  1 +  (1 - \rho^2) P_2 \right)  \ , \\
R_1  &\leq& \dfrac{1}{2} \displaystyle\min_{j\in[1:N]}  \log_2 \left(  1 + b_j^2 P_2 + P_1 + 2 b_j\rho \sqrt{ P_1 P_2} \right)   \ , \\
R_1 + R_2 &\leq& \dfrac{1}{2} \displaystyle\min_{j\in[1:N]} \biggl[  \log_2 \left(  1 + b_j^2 P_2 + P_1 + 2 b_j\rho \sqrt{ P_1 P_2} \right) \nonumber\\
&+& \dfrac{1}{2} \log^+_2\left( \dfrac{1 + (1 - \rho^2)P_2}{1 + (1 - \rho^2)P_2b_j^2}\right)   \biggr] \ , 
\end{IEEEeqnarray} 
where $\log_2^+ (x) \equiv  \max\{0, \log_2(x)\}$. 

Since $|b_j| > 1$ for all $j \in [1:N]$, then the outer bound becomes equal to:
\begin{IEEEeqnarray}{rCl}
R_2 &\leq&  \dfrac{1}{2} \log_2 \left(  1 + (1 - \rho^2) P_2 \right) \ ,  \\
R_1 + R_2 &\leq& \dfrac{1}{2} \displaystyle\min_{j\in[1:N]}   \log_2 \left(  1 + b_j^2 P_2 + P_1 + 2 b_j\rho \sqrt{ P_1 P_2} \right)  \  ,  
\end{IEEEeqnarray} 
which proves our claim. 
 \subsection{Weak Interference}
 The achieavability follows from the achievable rate region: 
\begin{equation}
\left\{\begin{array}{rcl}
R_1 &\leq& \displaystyle\min_{j \in[1:N]} I(X_1 U;  Y_j)  \ , \\
R_2 &\leq&  I(V;Z) - I(V; X_1U)   \  , 
\end{array}\right.
\end{equation} 
obtained through the inner bound~\eqref{CDcoding} by letting $Q_1 = Q = \emptyset$ and considering only one of the resulting corner points. 

The optimal coding scheme is then to let: 
\begin{eqnarray}
X_2 &=& X_u + X_v  \quad , \quad    X_v \sim \mathcal{N}\left(0,\eta P_2 \right) \quad , \quad    X_u  \sim   \mathcal{N}\left(0,(1-\eta) P_2 \right) \ , \\ 
  U &=& X_u \quad , \quad  X_1  \sim  \mathcal{N}(0,P_1) \ , \\
   (X_1, X_u) &\sim&  \mathcal{N} 
   \left(\mathbf{0},  \Sigma_{X_1X_u} \right) \ , \   \Sigma_{X_1X_u}=\left[ \begin{array}{cc}
P_1 &  \rho \sqrt{ P_1 (1 -\eta)P_2} \\ 
 \rho \sqrt{ P_1 (1 -\eta)P_2} & (1 - \eta) P_2 
\end{array} \right]\\ 
V &=& X_v + \gamma (X_u + a X_1) \ , 
\end{eqnarray}
where $|\rho| \leq 1$ and $\gamma$ is the optimal Dirty-Paper Coding parameter to precode against the interference $X_u + a X_1$ seen at user $Z$. 

Thus, we obtain: 
\begin{IEEEeqnarray}{rCl}
I(X_u X_1;Y_j) &=& h(Y_j) - h(Y_j | X_u X_1 ) \\
&=&  \dfrac{1}{2} \log_2 \left( \dfrac{1 + b_j^2 P_2 + P_1 + 2 b_j \rho \sqrt{(1- \eta ) P_1 P_2}}{ 1 + b_j^2 \eta P_2} \right)  \ , 
\end{IEEEeqnarray}
and 
\begin{equation}
I(V;Z) - I(V;UX_1) = I( X_v;Z  | X_1 X_u) = \dfrac{1}{2} \log_2 \left(  1 + \eta P_2 \right) \ . 
\end{equation}
\textit{N.B: It is because we can apply DPC techniques for the Gaussian case that we are able to relax the constraint of \emph{very weak interference} to only weak interference. In the general case, it is not possible for user $Z$ to decode interference unless its resulting sum-rate is satisfied, i.e., $R_1 + R_2 \leq I(X_1 X_2; Z)$. 
}

The parameter $\rho$ cannot be optimized for each instance of the primary channels $Y_j$ since the $b_j$'s are not all compulsorily equal (in sign and module), leading us to the $\max \min $ expression. 

Hereafter, the converse proof. Let us start by writing 
\begin{IEEEeqnarray}{rCl}
 nR_2 &=& H(W_2) \\
 &\overset{(a)}{\leq}& I(W_2; Z^n ) + n \epsilon_n \\
 &\overset{(b)}{\leq}& I(W_2; Z^n | W_1 X_1^n) + n \epsilon_n \\
 &\overset{(c)}{\leq}& I(X_2^n; Z^n | W_1 X_1^n) + n \epsilon_n  \ , 
\end{IEEEeqnarray} 
where $(a)$ is a consequence of Fano's inequality and $(b)$ follows from the fact that $W_2$ is independent of both $W_1$ and $X_1^n$, and $(c)$ results from the fact that the following Markov Chain holds $W_2 \mkv (X_2^n, X_1^n, W_1) \mkv Z^n$.

 Then, let $j \in [1:N]$, we have that: 
\begin{IEEEeqnarray}{rCl}
 nR_1 &=& H(W_1) \\
 &\leq& I(W_1; Y_j^n) + n \epsilon_n \\ 
 &\leq& I(W_1 X_1^n; Y_j^n) + n \epsilon_n \  .  
\end{IEEEeqnarray} 
Next, we bound the two resulting rates. Since,
 \begin{equation}
 \dfrac{n}{2}\log_2(2 \pi e) \leq  h(Z^n | W_1 X_1^n) \leq \dfrac{n}{2} \log_2( P_2 + 1) + \dfrac{n}{2}\log_2(2 \pi e) \ , 
 \end{equation}
 then 
  \begin{equation}
 \exists \,\eta \in [0:1] \quad \text{such that} \quad  h(Z^n | W_1 X_1^n) = \dfrac{n}{2} \log_2( \eta P_2 + 1) + \dfrac{n}{2}\log_2(2 \pi e) \ , 
 \end{equation}
 and since, 
 \begin{equation}
  h(Z^n | X_2^n X_1^n) =\dfrac{n}{2}\log_2(2 \pi e) \ , 
 \end{equation}
 we can conclude that
 \begin{equation}
  R_2 \leq \dfrac{1}{n} I(X_2^n ; Z^n | W_1 X^n_1) =\dfrac{1}{2} \log_2 ( \eta P_2 + 1) \ .
 \end{equation}
 Next, note that, with an abuse of notations: 
 \begin{equation}
  Y^n_j | X^n_1 = b_j X_2^n + n_2^n = b_j ( Z^n | X^n_1 )  +  b_j  \tilde{N}_2^n   \ , 
 \end{equation}
 where $ \tilde{N}_2 \sim \mathcal{N}\left(0,b_j^{-2} - 1\right)$.
 
 Thus, we can write by the $n$-letter conditional EPI that: 
  \begin{eqnarray}
 h(Y^n_j |W_1 X^n_1 ) &\geq& \dfrac{n}{2} \log_2\left( 2^{\frac{2}{n}h(b_j Z^n | W_1X^n_1 )}+  2^{\frac{2}{n}h( b_j \tilde{N}_2^n)} \right) \\
 &=& \dfrac{n}{2}\log_2\left[ b_j^2 2^{\frac{2}{n}h( Z^n | W_1X^n_1 )}+  2\pi e(1 - b_j^2)\right]\\
 &=& \dfrac{n}{2} \log_2\left[ b_j^2  ( \eta P_2 + 1) +   1 - b_j^2\right]+ \dfrac{n}{2}\log_2(2 \pi e) \\
 &=& \dfrac{n}{2} \log_2\left[ b_j^2  \eta P_2  +   1 \right] + \dfrac{n}{2}\log_2(2 \pi e)  \ . 
\end{eqnarray}
In addition, letting besides $(X_1,X_2)$ to have the following covariance matrix: 
\begin{equation}
 K = \left[ \begin{array}{cc}
P_1 &  \rho_{12} \sqrt{ P_1 P_2} \\ 
  \rho_{12} \sqrt{ P_1 P_2} &  P_2 
\end{array} \right] \ , 
\end{equation} 
we can combine with:  
\begin{eqnarray}
 \dfrac{n}{2}\log_2\left[ 2 \pi e (b_j^2  \eta P_2  +   1 )\right] & \leq&  h(Y^n_j |W_1 X^n_1 )  \\
 &\leq& h(Y^n_j |X^n_1)\leq \dfrac{n}{2} \log_2 \big[2 \pi e(  b_j^2(1- \rho_{12}^2)P_2+1) \big] \ , 
\end{eqnarray}
thus, 
\begin{equation}
 \eta \leq  1- \rho_{12}^2  \ .  
\end{equation}
Finally, we let $\rho \in [-1:1]$ such that
\begin{equation}
 \rho = \dfrac{\rho_{12}}{\sqrt{1- \eta}} \ , 
\end{equation}
to obtain: 
\begin{equation}
 R_1 \leq \dfrac{1}{2} \log_2\left(1 +  b_j^2  \eta P_2  +  P_1 + 2 b_j \rho\sqrt{P_1(1-\eta)P_2}  \right) - \dfrac{1}{2} \log_2 \left( b_j^2  \eta P_2  +   1 \right) \ , 
\end{equation}
which concludes the proof.
 \subsection{Mixed Weak/Very Strong Interference}
 The proof of the achievability follows by evaluating the rate region given by: 
   \begin{equation}
   \left\{ \begin{array}{rcl}
   R_1 &\leq& \displaystyle \min_{j \in \mathcal{W}} I(U X_1;Y_j) \ ,  \\
   R_2 &\leq& I(X_2; Z | UX_1) \ ,  \\
   R_1 + R_2 & \leq &  \displaystyle\min_{j \in \mathcal{S}} I(X_1 X_2;Y_j) \ ,
   \end{array}\right.
   \end{equation}
   with the following coding scheme:
   \begin{eqnarray}
X_2 &=& X_u + X_v  \quad , \quad    X_v \sim \mathcal{N}\left(0,\eta P_2 \right) \quad , \quad    X_u  \sim   \mathcal{N}\left(0,(1-\eta) P_2 \right) \ ,   \\ 
  U &=& X_u \quad , \quad  X_1  \sim  \mathcal{N}(0,P_1)\ , \\
   (X_1, X_u) &\sim&  \mathcal{N} 
   \left(\mathbf{0},  \Sigma_{X_1X_u} \right)\ , \ \Sigma_{X_1X_u} =\left[ \begin{array}{cc}
P_1 &  \rho \sqrt{ P_1 (1 -\eta)P_2} \\ 
 \rho \sqrt{ P_1 (1 -\eta)P_2} & (1 - \eta) P_2 
\end{array} \right]
\end{eqnarray}
where $|\rho| \leq 1$. Then, we obtain: 
\begin{eqnarray}
I(X_u X_1;Y_j) &=& h(Y_j) - h(Y_j | X_u X_1 ) \\
&=&  \dfrac{1}{2} \log_2 \left( \dfrac{1 + b_j^2 P_2 + P_1 + 2 b_j \rho \sqrt{(1- \eta ) P_1 P_2}}{ 1 + b_j^2 \eta P_2} \right)  \ , 
\end{eqnarray}
and 
\begin{equation}
I( X_2;Z  | X_1 X_u) = \dfrac{1}{2} \log_2 \left(  1 + \eta P_2 \right) \ .  
\end{equation}
Finally, 
\begin{equation}
I( X_2  X_1 ; Y_j) = \dfrac{1}{2} \log_2 \left(1 + b_j^2 P_2 + P_1 + 2 b_j \rho \sqrt{(1- \eta ) P_1 P_2} \right) \ .  
\end{equation}
   As for the outer bound, we can similarly to the weak and strong interference case, write that: 
   \begin{eqnarray}
 nR_1 &\leq& \min_{j \in \mathcal{W}} I(W_1 X_1^n; Y_j^n) + n \epsilon_n \ , \\
  nR_2 &\leq&  I(X_2^n; Z^n | W_1  X_1^n) + n \epsilon_n \ , \\
  n(R_1 + R_2) &\leq&  \min_{j \in \mathcal{S}} I(X_1^n X_2^n; Y_j^n) + n \epsilon_n   \ . 
\end{eqnarray} 
In the same fashion again, define $\eta \in [0:1]$ such that: 
\begin{equation} 
 h(Z^n | W_1 X_1^n) = \dfrac{n}{2} \log_2( \eta P_2 + 1) + \dfrac{n}{2}\log_2(2 \pi e) \ .
\end{equation}
We can show that for all $j \in \mathcal{W}$, 
\begin{equation} 
 h(Y_j^n | W_1 X_1^n) \geq  \dfrac{n}{2} \log_2( \eta b_j^2 P_2 + 1) + \dfrac{n}{2}\log_2(2 \pi e) \ . 
\end{equation}
As for $h(Y_j^n) $ with $j \in [1:N]$, it is maximized when $(X_1, X_2) $ follow a joint Gaussian distribution with covariance matrix: 
\begin{equation}
 K = \left[ \begin{array}{cc}
P_1 &  \rho_{12} \sqrt{ P_1 P_2} \\ 
  \rho_{12} \sqrt{ P_1 P_2} &  P_2 
\end{array} \right] \ , 
\end{equation}  where $\rho_{12} \in [-1:1]$.

Now, it can be noticed that $\rho_{12}$ has to satisfy the inequality: 
\begin{equation}
 \eta \leq  1- \rho_{12}^2  \ .  
\end{equation}
Let $\rho \in [-1:1]$ such that
\begin{equation}
 \rho = \dfrac{\rho_{12}}{\sqrt{1- \eta}} \ ,  
\end{equation}
which yields an outer bound equasl to the claimed capacity region~\eqref{CapacityMixed}. 

\subsection{Coherent and non-coherent Interference}
Since the three claims have similar proofs, we only show the first one. The first part of the proof is trivial since the optimal $\rho$ is obtained by $+1$ if all channel coefficients $b_j$ are positive, and $-1$ otherwise.  The second half of the claim can be proved by letting $C_j$ be the capacity region of the CIFC $(Z,Y_j)$ in weak interference. Assume that all $b_j$'s are positive. We have that: 
 \begin{equation}
\mathcal{C}_j(\eta)\ : \ \left\{\begin{array}{rcl}
R_1 &\leq&  R_{1,j}(\eta) \equiv  \dfrac{1}{2}  \ \log_2 \left( \dfrac{1 + b_j^2 P_2 + P_1 + 2  b_j \sqrt{ (1 - \eta) P_1 P_2}}{ 1 + b_j^2 \eta P_2} \right)  \ , \\
R_2 &\leq&   R_{2}(\eta) \equiv  \dfrac{1}{2} \log_2 \left( 1 + \eta P_2  \right) \ . 
\end{array}\right.
\end{equation}
We want to show that: 
\begin{equation}
\bigcap_{j=1}^N \bigcup_{\eta\in[0:1]} \mathcal{C}_j(\eta) =  \bigcup_{\eta\in[0:1]} \mathcal{C}(\eta) \ .
\end{equation}
We have that for all $j \in[1:N]$, $\mathcal{C}(\eta) \subset \mathcal{C}_j(\eta)$, thus it is easy to prove the first inclusion:  
\begin{equation}
  \bigcup_{\eta\in[0:1]} \mathcal{C}(\eta) \subset\bigcap_{j=1}^N \bigcup_{\eta\in[0:1]} \mathcal{C}_j(\eta)   \ . 
\end{equation}
Now, to show the other inclusion, let $(R_1,R_2)$ be a rate pair in $\bigcap_{j=1}^N \bigcup_{\eta\in[0:1]} \mathcal{C}_j(\eta)$. We need to show that $(R_1,R_2)$ lies in $\bigcup_{\eta\in[0:1]} \mathcal{C}(\eta)$. Let $\eta_1\dots\eta_N$ be N parameters such that: 
\begin{equation}
   \left\{\begin{array}{rcl}
   R_1 &\leq& \displaystyle \min_{j\in[1:N]} R_{1,j}(\eta_j)  \ ,  \\
   R_2 &\leq& \displaystyle  \min_{j\in[1:N]} R_{2}(\eta_j) \ .  
   \end{array}\right.
  \end{equation}  
Note that, $R_2$ is increasing in $\eta$ while $R_{1,j}$ is decreasing in $\eta$. 
Thus, we have
\begin{equation}
 R_2  \leq  \displaystyle \min_{j\in[1:N]} R_{2}(\eta_j) =  R_{2}\left( \min_{j\in[1:N]} \eta_j\right)  \ . 
\end{equation} 
And hence, since $R_{1,j}$ is decreasing in $\eta$, then: 
\begin{equation}
 R_1  \leq  \displaystyle \min_{j\in[1:N]} R_{1,j}( \eta_j) \leq \displaystyle \displaystyle \min_{j\in[1:N]} R_{1,j}\left( \displaystyle \min_{j\in[1:N]} \eta_j\right) \ , 
\end{equation}
thus, setting $\eta \equiv\displaystyle\min_{j\in[1:N]} \eta_j$ allows $(R_1,R_2)$ to lie in the region $\bigcup_{\eta\in[0:1]} \mathcal{C}(\eta)$.
 
 \section{Multi-Secondary Very Weak Interference Capaacity region}\label{App-VWI-MultiSecondary}  
    
From Fano's inequality, we can write that: 
\begin{IEEEeqnarray}{rCl}
n \, (R_1 - \epsilon_n) &\leq& I(W_1; Y^n) \\
&\leq& I(W_1X_1^n; Y^n) \\ 
&\leq&    \sum_{i=1}^n I(W_1 X_{1,i} X_{1,<i>} Y^{i-1} ; Y_{i} )  \ , 
\end{IEEEeqnarray}
where $ X_{1,<i>} \equiv (X_1^{i-1},X_{1,i+1}^n)$. 

Similarly, for the other rate, let $k \in [1:M]$: 
\begin{IEEEeqnarray}{rCl}
 n \, (R_2 - \epsilon_n)   &\leq& I(X_2^n; Z_k^n|W_1X_1^n)  \\
&=& \sum_{i=1}^n I(X_{2,i}; Z_{k,i}| W_1  X_{1,i} X_{1,<i>} Z^{i-1}_k) \\
&=& \sum_{i=1}^n \bigl[ I(X_{2,i}; Z_{k,i}| W_1  X_{1,i} X_{1,<i>}) \nonumber  \\
 && \qquad - I(Z^{i-1}_k; Z_{k,i}| W_1  X_{1,i} X_{1,<i>} ) \bigr] \\
&\overset{(a)}{\leq}& \sum_{i=1}^n \bigl[ I(X_{2,i}; Z_{k,i}| W_1  X_{1,i} X_{1,<i>})\nonumber  \\
 && \qquad  - I(Y^{i-1}; Z_{k,i}| W_1  X_{1,i} X_{1,<i>} ) \bigr]\\
&=& \sum_{i=1}^n I(X_{2,i}; Z_{k,i}| W_1  X_{1,i} X_{1,<i>} Y^{i-1})   \ , 
\end{IEEEeqnarray}
where $(a)$ is a result of \eqref{EquWeakIFC} and is proved as follows. 
Let $i \in [1:n]$, we have that: 
      \begin{IEEEeqnarray}{rCl} 
 I( W_1 X^n_1 Z_k^{i-1};  Y_{i} ) &-&  I( W_1 X^n_1 Y^{i-1};  Y_{i} ) \nonumber \\
    &=& I(Z_k^{i-1};  Y_{i} | W_1 X^n_1  )  -  I(Y^{i-1} ;  Y_{i} |  W_1 X^n_1) \nonumber  \\
    &=& \sum_{r= 1}^{i-1}  \Bigl[ I( Y^{r-1} Z_{k,r}^{i-1} ; Y_{i} |   W_1 X^n_1 ) - I( Y^{r} Z_{k,r+1}^{i-1} ; Y_{i} |   W_1 X^n_1 ) \Bigr] \nonumber \\
    &=& \sum_{r=1 }^{i-1} \Bigl[ I( Z_{k,r}  ; Y_{i} | X_{1,r}  Y^{r-1} Z_{k,r+1}^{i-1}  W_1  X^n_{1,r+1} X_1^{r-1} ) \nonumber   \\ 
 	 &- &I( Y_{r}  ; Y_{i} |   X_{1,r} Y^{r-1} Z_{k,r+1}^{i-1}  W_1  X^n_{1,r+1} X_1^{r-1})   \Bigr] \overset{(a)}{\geq}  0 \, \nonumber  . 
   \end{IEEEeqnarray}
   The weak interference condition implies that for  $r\in [1:i-1]$:
    \begin{equation}
 I(U ; Y_{r} | V X_{1,r})  \leq I(U ; Z_{k,r} | V  X_{1,r}) \ , 
\end{equation}
   where we let $U  = Y_{i}$  and $V \equiv (Y^{r-1}, Z_{k,r+1}^{i-1}, W_1, X^n_{1,<r>} )$.  Thus, letting $U_i  \equiv( W_1,X_{1,<i>},Y^{i-1})$ completes the proof. 
 
\bibliographystyle{IEEEtran}
\bibliography{BiblioMeryemBenammar} 

\end{document}